\documentclass[10pt]{revtex4}
\usepackage{amsmath,amssymb}
\usepackage{graphicx,epstopdf}
\usepackage[usenames,dvipsnames,svgnames]{xcolor}
\usepackage{hyperref}
\usepackage{array}
\usepackage{multirow}
\usepackage{blindtext}
\usepackage{relsize}

\begin{document}
	\title{A new shape function and some specific wormhole solutions in braneworld scenario}
	\author{Bikram Ghosh\footnote {bikramghosh13@gmail.com}}
	\author{Saugata Mitra\footnote { saugatamitra20@gmail.com}}
	\affiliation{Department of Mathematics, Ramakrishna Mission Vidyamandira, Howrah-711202, West Bengal, India}

	%%%%%%%%%%%%%%%%%%%%%%%%%%%%%%%%%%%%%%%%%%%%%%%%%%%%%%%%%%%%%%%%%%%%%%%%%%%%%%%%%%%%%%%%%%%%%%%%%%%%%%%%%%%%%%%%%%%%%%%%%%%%
	
	\begin{abstract}
	Considering a energy density of the form $\rho=q\left({\frac{r}{r_0}}\right)^{-n}$( where $q$ is an arbitrary positive constant with dimension of energy density and $n>0$), a shape function is obtained by using field equations of braneworld gravity theory in this paper. Under isotropic scenario wormhole solutions are obtained considering six different redshift functions along with the obtained new shape function. For anisotropic case wormhole solutions are obtained under the consideration of five different shape functions along with the redshift function $\phi=\beta ln(\frac{r}{r_0})$, where $\beta$ is an arbitrary constant. In each case all the energy conditions are examined and it is found that for some cases all energy conditions are satisfied in the vicinity of the wormhole throat and for the rest cases all energy conditions are satisfied except strong energy condition.
		
	\end{abstract}
	\maketitle
		
	\section{Introduction}
	\label{int}
	In the context of string theory\cite{b1}--\cite{b2}, the brane ({\it i.e.,} the observed four dimensional world) is considered as a domain wall in the five dimensional space-time. Geometrically, brane\cite{b3}--\cite{b4} is considered as a four dimensional hypersurface embedded in a five dimensional bulk (for a review see\cite{b5}--\cite{b6}). In Randall-Sundrum type-II\cite{b4} brane model, our universe ({\it i.e.,} 3 brane) is embedded in a five-dimensional bulk with extra dimension extending to infinity in either side of the brane. In this model all standard model fields are confined to the brane while gravity can propagate in the surrounding bulk. Starting from the five-dimensional Einstein field equations in the bulk and using Gauss and Codazzi equations Shiromizu et al.\cite{b7} derived the modified four-dimensional Einstein equations on the brane. The above modified Einstein field equations has two extra terms in the r.h.s-- one correction term is quadratic in the energy-momentum tensor on the brane and is termed as local bulk effect while the non-local bulk effect is the electric part of the $5D$ Weyl tensor. Hence the effective Einstein equations on the brane can be written as \cite{r24}--\cite{r15}
		\begin{equation}\label{eq2}
	G_{\mu\nu}=-\Lambda g_{\mu\nu}+k^2(T_{\mu\nu}+\frac{6}{\lambda}P_{\mu\nu})-\xi_{\mu\nu},
	\end{equation} where the terms on the r.h.s of the above modified Einstein equation are the followings: $\Lambda=\frac{1}{2}(\Lambda_{5}+k^2\lambda)$, $k^2=\frac{\lambda k_5^2}{6}$ ,
		\begin{equation}
	P_{\mu\nu}=\frac{1}{8}\bigg[\frac{2}{3}TT_{\mu\nu}-2T_{\mu\psi}T_\nu^\psi+g_{\mu\nu}\biggl\{T_{\psi\tau }T^{\psi\tau}-\frac{1}{3}T^2\biggr\}\bigg],
	\end{equation}
	and 
	 $\xi_{\mu\nu}=\delta_{\mu}^A\delta_{\nu}^CW_{ABCD}\eta^B\eta^D$.
	 Here $k^2$ and $k_5^2$ are gravitational coupling constants on brane and bulk respectively, $\Lambda$ and $\Lambda_{5}$ are the cosmological constants on the brane and bulk respectively, $\lambda$ is the brane tension, $T_{\mu\nu}$ is the energy momentum tensor for the matter on the brane
%	  with traceless \text{\it i.e.,
%	  	 $\xi_\mu^\mu=0$}, is the projection of the 5-dimensional Weyl-tensor $W_{ABCD}$ onto the brane , where $\eta^A$ is the unit normal to the brane
  	  (with $T=T_{\mu\nu}g^{\mu\nu}$ is the trace of the energy momentum tensor), $P_{\mu\nu}$ is the local correction term, $\xi_{\mu\nu}$ is the (trace less {\it i.e.,} $\xi^\mu_\mu=0$) non-local bulk effect(which is the projection of the $5D$ dimensional Weyl tensor $W_{ABCD}$) and $\eta^a$ is the unit normal to the brane.
	  \par 
	  Wormhole is a hypothetical geometric structure in space-time \cite{r1}--\cite{r3}. It can be considered as tunnels in the space-time topology. In fact this topological structure connects two distant parts of a single universe or even of different universes. The wormhole geometry is characterized by the solution of the Einstein equations with exotic matter (that violates the null energy condition) at least in the vicinity of the wormhole throat\cite{r6}--\cite{r8}. In spherically symmetric space-time, traversable wormhole is most interesting and physically viable and such type of wormhole was initiated by the pioneering work of Morris and Throne\cite{r4}--\cite{r5}. The restriction in the geometry of the space-time due to the traversability nature of the wormhole is that the redshift function should not have any horizon or it is desirable to have a given asymptotic form for both the redshift and shape function. Usually, in the literature wormhole solutions are constructed knowing a priori the desired form of the redshift function and the shape function and Einstein field equations determine the matter part for the wormhole geometry. 
	\par 

%	\par 
%	The notion of wormhole came to light by Flamm\cite{r1} in 1916. Wormholes are topological tunnel like hypothetical structures which connects two distant regions in same space-time or it connects two distant universes. Einstein and Rosen \cite{r2} study wormhole's nature and at that time it was known as ``Einstein-Rosen Bridge". Misner and Wheeler\cite{r3} first named ``Einstein-Rosen bridge" as ``wormhole" in 1957. The study of wormhole, accelerate after the work of Morris and Thorne\cite{r4}, where the presence of traversable wormholes is shown. By the word `traversable' it is meant that a human could travel through the wormhole within a reasonable amount of time\cite{r5}. Visser\cite{r6} and from other studies\cite{r4},\cite{r7} it was observed that near the wormhole throat, matter must violate null energy condition. This type of  matter is known as exotic matter and they are necessary to form a wormhole in Einstein gravity theory.
%\par
%	However in modified gravity theory, violation of energy conditions is not necessary. In $f(R)$ gravity theory it is observed that by choosing suitable form of $f(R)$ and other components not only null energy condition satisfied, weak and dominant energy conditions are also satisfied\cite{r5}, \cite{r8}. In $f(R,T)$ gravity theory by choosing suitable form of $f(R,T)$ and other components null and weak energy conditions are satisfied\cite{r9}-\cite{r10}. In Rastal gtravity theory non-violation of null and weak energy conditions are also observed\cite{r11}.
	\par
	Braneworld gravity\cite{r12} is an important model of the universe in modified gravity theory and obtainig wormhole solutions in this gravity is attracted to researchers\cite{r13}-\cite{r17}, where wormhole solutions are obtained under different considerations. In \cite{r24},  wormhole solutions are obtained by equating $R$ with zero. Inflating wormhole solutions are obtained in\cite{r18}. In the literature there are many examples of wormhole solutions where null energy condition is satisfied\cite{r20}-\cite{r23}.
	\par 
	In this paper, a new shape function is found by considering a particular form of energy density and the main motivation is to examine whether all energy conditions are satisfied or not in the wormhole solutions which are obtained by considering isotropic fluid and  using the new shape function along with the following redshift functions: $\phi(r)=\text{constant}$; $\phi(r)=\beta ln\left(\frac{r}{r_0}\right)$, $\beta$ is a arbitrary constant; $\phi(r)=\frac{1}{r}$; $\phi(r)=ln\frac{\sqrt{\gamma^2+r^2}}{r}$, $\gamma$ is a arbitrary constant; $\phi(r)=e^{-\frac{r_0}{r}}$; $\phi(r)=e^{-\frac{r_0}{r}-\frac{r_0^2}{r^2}}$\cite{r25}-\cite{r26}. Other wormhole solutions for anisotropic fluid using shape functions: $b(r)=\frac{r_0^n}{r^{n-1}}$,($n>0$); $b(r)=\frac{r}{1+ln(1+r-r_0)}$, ($0<r_0<1$); $b(r)=\frac{r}{1+r-r_0},~(0<r_0<1)$; $b(r)=r_0\biggl\{1+\gamma^2\biggl(1-\frac{r_0}{r}\biggr)\biggr\}$, ($\gamma^2\in(0,1)$)\cite{r19} and $b(r)=re^{\frac{2\sigma}{\delta}(r^\delta-r_0^\delta)}$ ($\sigma<0$ and $\delta>0$)\cite{r25}-\cite{r26}  and redshift function $\phi(r)$=$\beta ln(\frac{r}{r_0})$ are obtained and energy conditions are also examined in this cases.\\ The paper is arranged as follows : in section \ref{secii}, the necessary field equations on Brane-world are discussed. In section \ref{ec} energy conditions of wormholes are mentioned. A wormhole shape function is obtained in section \ref{seciii}. Energy conditions are examined in section \ref{wsi} for isotropic fluid. For anisotropic fluid wormhole solutions are obtained, energy conditions are examined in section \ref{wsa}. Lastly, we discuss our overall observations in section \ref{secv}.   
	\section{Field equations on the brane}
	\label{secii}
	The space-time metric representing a spherically symmetric and static wormhole is given by \cite{r4}
	\begin{equation}\label{eq1}
	ds^2=-e^{2\phi(r)}dt^2+\frac{dr^2}{1-\frac{b(r)}{r}}+r^2(d\theta^2+{\sin^2\theta}d\phi^2) , \end{equation}
	where $\phi(r)$ and $b(r)$ are arbitrary function of radial co-ordinate $`r$', termed as redshift function and shape function respectively. The former is related to gravitational redshift, while the latter determines the shape of the wormhole. The event horizon should be absent for traversable wormhole, for which we require $e^{2\phi(r)}\neq0$ {\it i.e.,} $\phi(r)$ should be finite everywhere. Now for the existence of wormhole, the shape function $b(r)$ is restricted as follows $(i)$ $b(r_0)=r_0$ ($r_0$ is the location of the throat), $(ii)$ $ \frac{b(r)}{r}<1 $ for $r>r_0$, $(iii)$ $\frac{b(r)}{r}\rightarrow0$ and $\phi(r)\rightarrow\phi_0$(constant) as $r\rightarrow\infty$ (asymptotically flatness property\cite{r4}-\cite{r5}), $(iv)$ $
	\frac{b-b'r}{2b^2}>0$ is the condition for flare-out\cite{r4}.
	\par 
\par 
	The field equation on the brane can take the form
	\begin{equation}\label{eq5}
	G_{\mu\nu}=8\pi T_{\mu\nu}^{\text{eff}},
	\end{equation}
	with the total effective stress-energy tensor,
	\begin{equation}\label{eq6}
	T_{\mu\nu}^{\text{eff}}=T_{\mu\nu}-\frac{1}{8\pi}\xi_{\mu\nu}+\frac{6}{\lambda}P_{\mu\nu}\text{~~~~with~~} k^2=8 \pi.
	\end{equation}
	For simplicity we have considered $\Lambda=0$ on the brane and hence Einstein tensor components (with respect to an orthonormal reference frame) for the metric (\ref{eq1}) can be written as
	\begin{eqnarray}\label{eq7}
	G_{tt}&=&\frac{b'}{r^2},\\\label{eq8}
	G_{rr}&=&2\biggl(1-\frac{1}{r}\biggr)\frac{\phi'}{r}-\frac{b}{r^3},\\\label{eq9}
	G_{\theta\theta}=G_{\phi\phi}&=&\biggl(1-\frac{b}{r}\biggr)\biggl[\phi''+(\phi')^2+\frac{\phi'}{r}-\frac{b'r-b}{2r^2(r-b)}-\frac{b'r-b}{2r(r-b)}\phi'\biggr],
	\end{eqnarray}
	where the prime denotes the derivative with respect to the radial coordinate $`r$'.
		From (\ref{eq2}), the traceless property of the projected 5-dimensional Weyl-tensor gives\cite{r15} 
	\begin{equation}
	R=-k^2\biggl\{T+\frac{3}{2\lambda}\left(T_{\psi\tau }T^{\psi\tau}-\frac{1}{3}T^2\right)\biggr\}
	\end{equation}
	\text{\it i.e.,}
	\begin{equation}\label{eq10}
	R=-k^2\bigg[(-\rho+p_r+2p_t)+\frac{3}{2\lambda}\biggl\{\rho^2+p_r^2+2p_t^2-\frac{1}{3}\bigg(-\rho+p_r+2p_t\bigg)^2\biggr\}\bigg].
	\end{equation}
	Also from the definition of Ricci scalar ($R$)\cite{r30} we get,
	\begin{equation}\label{eq11.1}
	R=-2\biggl(1-\frac{b}{r}\biggr)\biggl[\phi''+(\phi')^2-\frac{b'}{r(r-b)}-\frac{b'r+3b-4r}{2r(r-b)}\phi'\biggr]
	\end{equation}
	and the result becomes $R|_{r=r_0}=\frac{2b_0'}{r_0^2}+\frac{(b_0'-1)\phi_0'}{r_0^2}$ at the throat. 
%	 For an isotropic fluid on the brane as $T_{\mu\nu}=$diag$(\rho,p,p,p)$ where $\rho$ is energy density, $p$ is isotropic radial pressure of the fluid. As a result, the nonlocal bulk effects contribute in the form an effective anisotropic fluid as 
%	\begin{equation}
%	\xi_{\mu\nu}=\text{diag}[\epsilon(r), \sigma_r(r), \sigma_t(r), \sigma_t(r)],
%	\end{equation}
%	and the traceless property of $\xi_{\mu\nu}$ gives
%	\begin{equation}\label{eq11}
%	-\epsilon+\sigma_r+2\sigma_t=0.
%	\end{equation}
%Now using equations (\ref{eq5}) and (\ref{eq6})	we get the effective field equations from (\ref{eq7})--(\ref{eq9}) as:
%	\begin{eqnarray}\label{eq12}
%	\frac{b^\prime}{r^2}&=&\rho (1+\frac{\rho}{2\lambda})-\frac{\epsilon}{8\pi},\\\label{eq13.2}	
%	\frac{2}{r}\left(1-\frac{b}{r} \right)\Phi^\prime -\frac{b}{r^3}&=&p(1+\frac{\rho}{\lambda})+\frac{\rho^2}{2\lambda}-\frac{\sigma_r}{8\pi},\\\label{eq14}
%	\left(1-\frac{b}{r} \right)\left[\Phi^{\prime \prime} +\Phi^\prime (\Phi^\prime +\frac{1}{r}) \right] -\frac{b^\prime r-b}{2r^2} (\Phi^\prime +\frac{1}{r})&=&p(1+\frac{\rho}{\lambda})+\frac{\rho^2}{2\lambda}-\frac{\sigma_t}{8\pi}.
%	\end{eqnarray}
		\section{Energy conditions}\label{ec}
	The null energy condition (NEC), weak energy condition (WEC), strong energy condition (SEC) and dominant energy conditions (DEC) are considered main energy conditions in the background of brane-world gravity. These will be investigated by the following inequalities depending on energy momentum tensor characterize the energy conditions as follows\cite{r29}-\cite{r30}:
	\begin{eqnarray}
	\label{eq13.1} &(I)& \text{NEC} : \rho+p_r\geq0, \rho+p_t\geq0\\\label{eq14.1}
	&(II)& \text{WEC} : \rho\geq0, \rho+p_r\geq0, \rho+p_t\geq0\\\label{eq15.1}
	&(III)& \text{SEC} : \rho+p_r\geq0, \rho+p_t\geq0, \rho+p_r+2p_t\geq0\\\label{eq16.1}
	&(IV)& \text{DEC} : \rho\geq0, \rho-|p_r|\geq0, \rho-|p_t|\geq0.
	\end{eqnarray}
	\par 
	For the isotropic fluid with radial pressure $p$, the above set becomes:
	
	\begin{eqnarray}
	\label{eq13.11} &(I)& \text{NEC} : \rho+p\geq0\\\label{eq14.11}
	&(II)& \text{WEC} : \rho\geq0, \rho+p\geq0\\\label{eq15.11}
	&(III)& \text{SEC} : \rho+p\geq0, \rho+3p\geq0\\\label{eq16.11}
	&(IV)& \text{DEC} : \rho\geq0, \rho-|p|\geq0
	.
	\end{eqnarray}
%	 Also we have cosidered all the components of projected Weyl tensor as zero.
%	From (\ref{eq2}), the traceless property of the projected 5-dimensional Weyl-tensor gives\cite{r15} 
%	\begin{equation}
%	R=-k^2\biggl\{T+\frac{3}{2\lambda}\left(T_{\psi\tau }T^{\psi\tau}-\frac{1}{3}T^2\right)\biggr\}
%	\end{equation}
%	\text{\it i.e.,}
%	\begin{equation}\label{eq10}
%	R=-k^2\bigg[(-\rho+p_r+2p_t)+\frac{3}{2\lambda}\biggl\{\rho^2+p_r^2+2p_t^2-\frac{1}{3}\bigg(-\rho+p_r+2p_t\bigg)^2\biggr\}\bigg].
%	\end{equation}
%	Also from the definition of Ricci scalar ($R$)\cite{r30} we get,
%	\begin{equation}\label{eq11}
%	R=-2\biggl(1-\frac{b}{r}\biggr)\biggl[\phi''+(\phi')^2-\frac{b'}{r(r-b)}-\frac{b'r+3b-4r}{2r(r-b)}\phi'\biggr]
%	\end{equation}
%	and the result becomes $R|_{r=r_0}=\frac{2b_0'}{r_0^2}+\frac{(b_0'-1)\phi_0'}{r_0^2}$ at the throat.
%
\section{Obtaining Shape function}\label{seciii}
 For an isotropic fluid on the brane, the energy momentum tensor can be written as $T^\mu_\nu=$diag$(-\rho,p,p,p)$\cite{r15} where $\rho$ is energy density, $p$ is isotropic radial pressure of the fluid. As a result, the nonlocal bulk effects contribute in the form an effective anisotropic fluid as 
\begin{equation}
\xi^\mu_\nu=\text{diag}[-\epsilon(r), \sigma_r(r), \sigma_t(r), \sigma_t(r)],
\end{equation}
and the traceless property of $\xi_{\mu\nu}$ gives
\begin{equation}\label{eq11}
-\epsilon+\sigma_r+2\sigma_t=0.
\end{equation}
Now using equations (\ref{eq5}) and (\ref{eq6})	we get the effective field equations from (\ref{eq7})--(\ref{eq9}) as\cite{r19}:
\begin{eqnarray}\label{eq12}
\frac{b^\prime}{r^2}&=&\rho (1+\frac{\rho}{2\lambda})-\frac{\epsilon}{8\pi},\\\label{eq13.2}	
\frac{2}{r}\left(1-\frac{b}{r} \right)\phi^\prime -\frac{b}{r^3}&=&p(1+\frac{\rho}{\lambda})+\frac{\rho^2}{2\lambda}-\frac{\sigma_r}{8\pi},\\\label{eq14}
\left(1-\frac{b}{r} \right)\left[\phi^{\prime \prime} +\phi^\prime (\phi^\prime +\frac{1}{r}) \right] -\frac{b^\prime r-b}{2r^2} (\phi^\prime +\frac{1}{r})&=&p(1+\frac{\rho}{\lambda})+\frac{\rho^2}{2\lambda}-\frac{\sigma_t}{8\pi}.
\end{eqnarray}
\par 
It is seen that in many wormhole theory literature the energy density takes the form $\rho=k r^{-s}$ where $k$ is a arbitrary positive constant, $s=1.5$ in \cite{r30} and $s=1.75$ in \cite{r31} . So let us take the form $\rho=q\left({\frac{r}{r_0}}\right)^{-n}$ for $n>0$(where $q$ will take care about the dimension of energy density) and  try to find shape function in this scenario.

Let us consider $\epsilon =0$. Then for the above energy density form $\rho=q\left({\frac{r}{r_0}}\right)^{-n}$ for $n>0$, the equation $(\ref{eq12})$ gives
\begin{eqnarray}
b^\prime &=& qr_0^nr^{2-n}+\frac{q^2r_0^{2n}}{2\lambda}r^{2-2n}. \\
\therefore~b&=& \frac{qr_0^n}{3-n}r^{3-n}+\frac{q^2r_0^{2n}}{2 (3-2n)\lambda }r^{3-2n}+A  ~(\text{where $A$ is an arbitrary constant}). \label{e1}
\end{eqnarray}
To be a shape function $b$ must obey the following properties:\\
\begin{eqnarray}
b(r_0)&=&r_0 ,\label{p1}\\
\frac{b}{r}&\leq &1 , \label{p2}\\
\frac{b-b^\prime r}{b^2}&>&0  \label{p3}.
\end{eqnarray}
Using $(\ref{p1})$, equation $(\ref{e1})$ reduces to
\begin{eqnarray}\label{b}
b(r)=\frac{qr_0^n}{3-n}r^{3-n}+\frac{q^2r_0^{2n}}{2 (3-2n)\lambda }r^{3-2n}+r_0-\frac{qr_0^3}{3-n}-\frac{q^2r_0^3}{2 (3-2n)\lambda }.
\end{eqnarray}
\begin{figure}[b]
	\centering
	% \hspace*{\fill}
	\begin{minipage}{.55\textwidth}
		\centering
		\includegraphics[width=.45\linewidth]{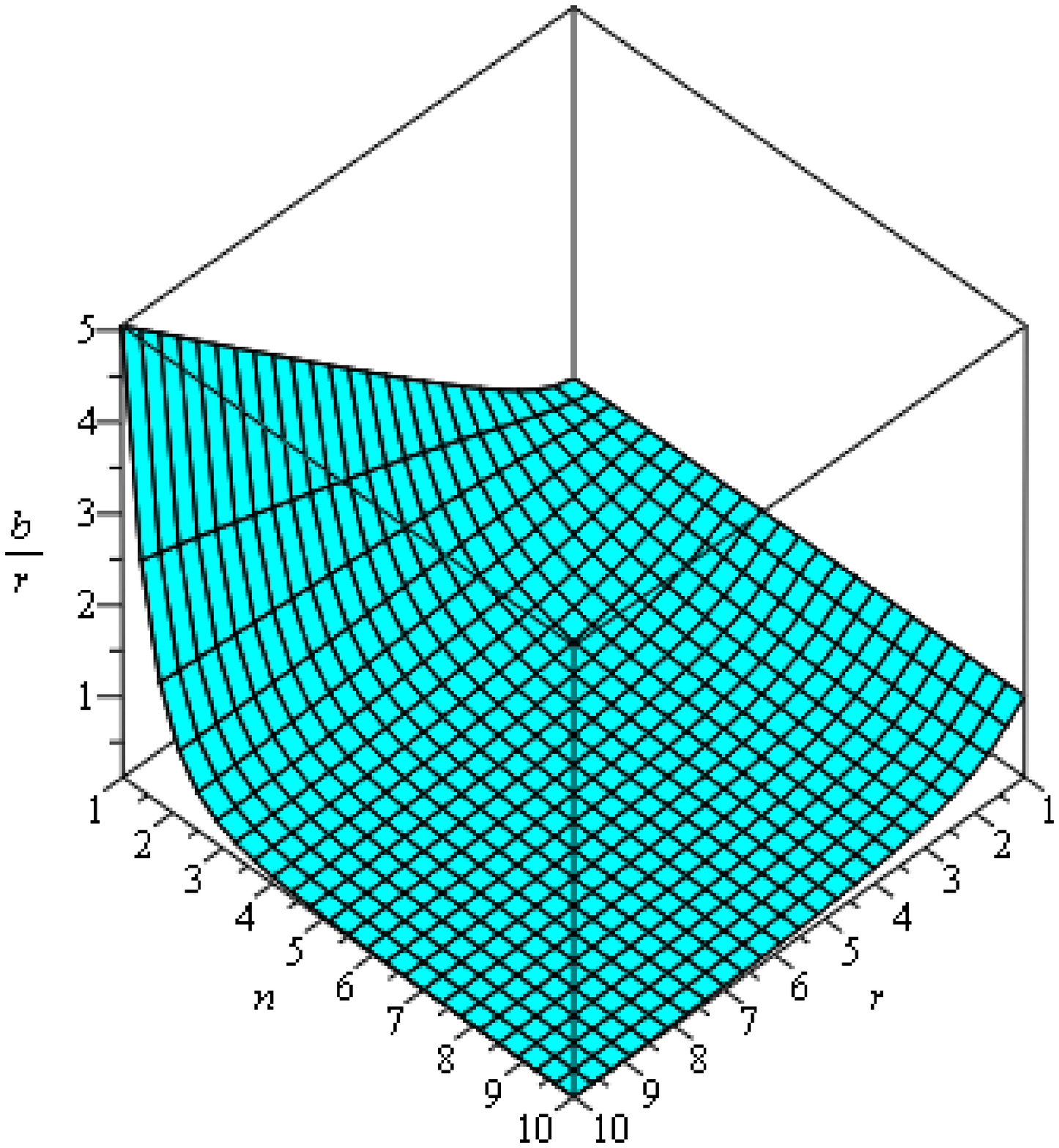}	
			\centering 1(A)
	\end{minipage}%%
	\begin{minipage}{.55\textwidth}
		\centering
		\includegraphics[width=.6\linewidth]{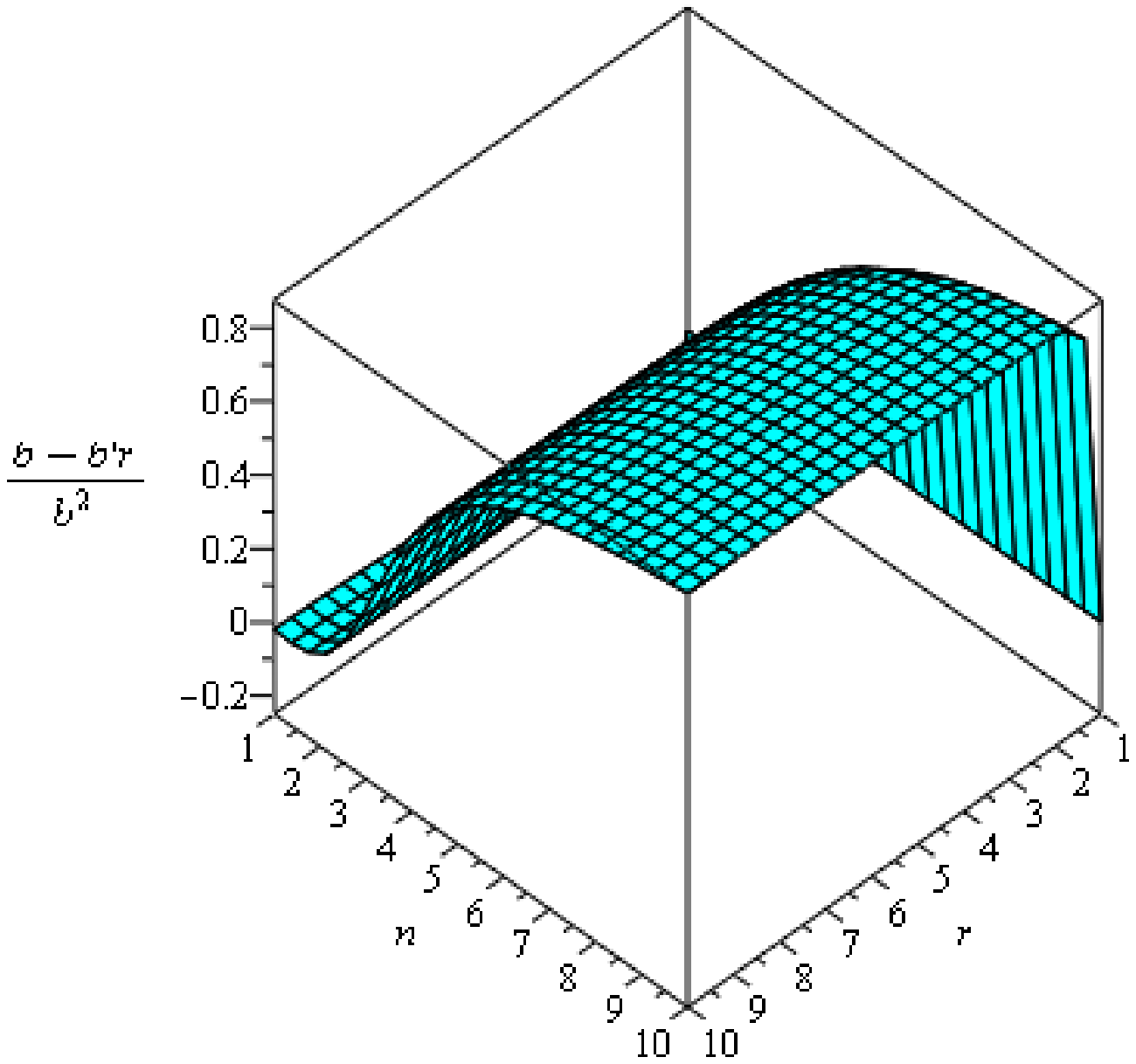}
			\centering 1(B)	
	\end{minipage}
	
	\caption{Behavior of $\frac{b(r)}{r}$(1(A)) and  $\frac{b-b'r}{b^2}$ (1(B))  with radial co-ordinate `$r$' and parameter $n$ have been plotted for obtained shape function(\ref{b}) when $1\leq n\leq10$ and $r$ ranges from $r_0$ to 10, $\lambda=10^4$, $q=1$ and $r_0=1$.}
	\label{fig1.1}
\end{figure}
\begin{figure}[!]
	\centering
	% \hspace*{\fill}
	\begin{minipage}{.55\textwidth}
		\centering

		\includegraphics[width=.6\linewidth]{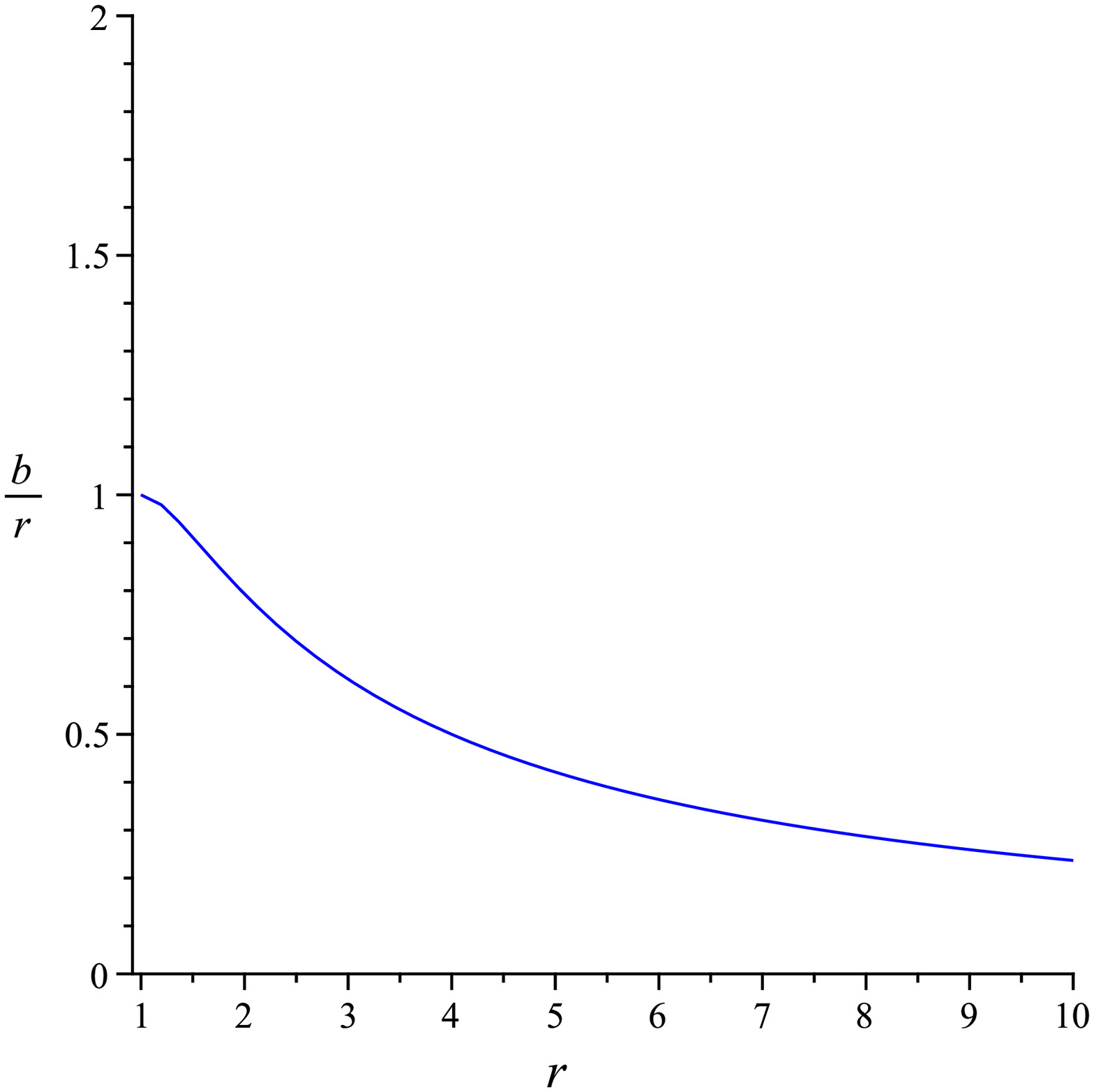}	
			\centering 2(A)
	\end{minipage}%%
	\begin{minipage}{.55\textwidth}
		%\centering
		\includegraphics[width=.6\linewidth]{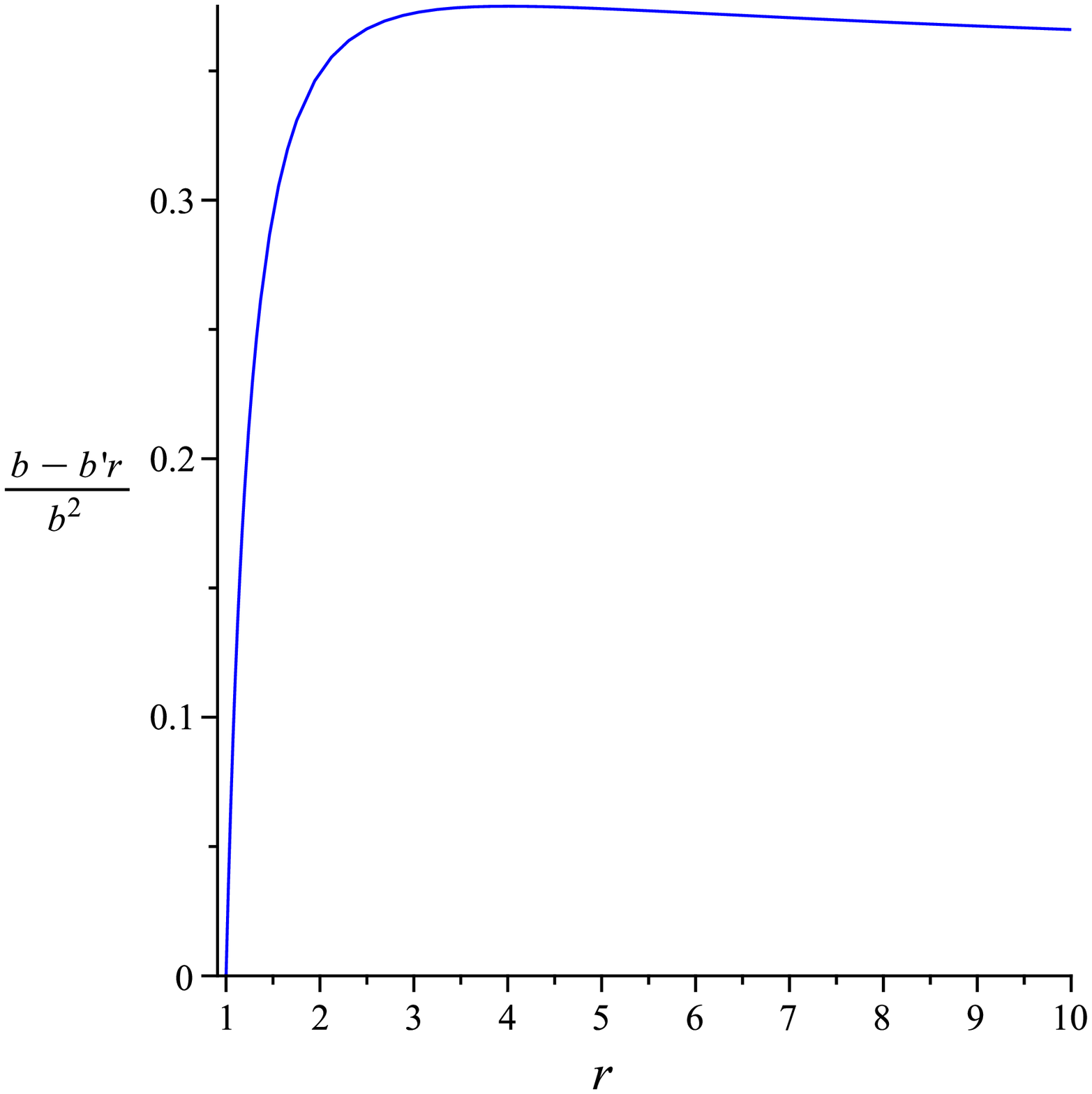}
			\centering 2(B)	
	\end{minipage}
	\caption{Behavior of $\frac{b(r)}{r}$(2(A)) and $\frac{b-b'r}{b^2}$ (2(B))
		 have been plotted for obtained shape function(\ref{b}) against $r$ when $n=3.5$, $\lambda=10^4$,$q=1$ and $r_0=1$.}
	\label{fig2}
\end{figure}

The figure(\ref{fig1.1}) shows that $b(r)$ is satisfying the other require conditions to be a shape function of a wormhole  for $n\in(3, 10)$ (a particular scenario is shown in figure\ref{fig2}). 
%The asymptotic nature of the obtained shape function is observed in the figure(\ref{fig2}(C)).
\section{Validation of energy condition of wormhole solutions corresponds to obtained shape function for isotropic fluid}
\label{wsi}
	\par
\begin{figure}
			\centering
		\includegraphics[width=.6\linewidth]{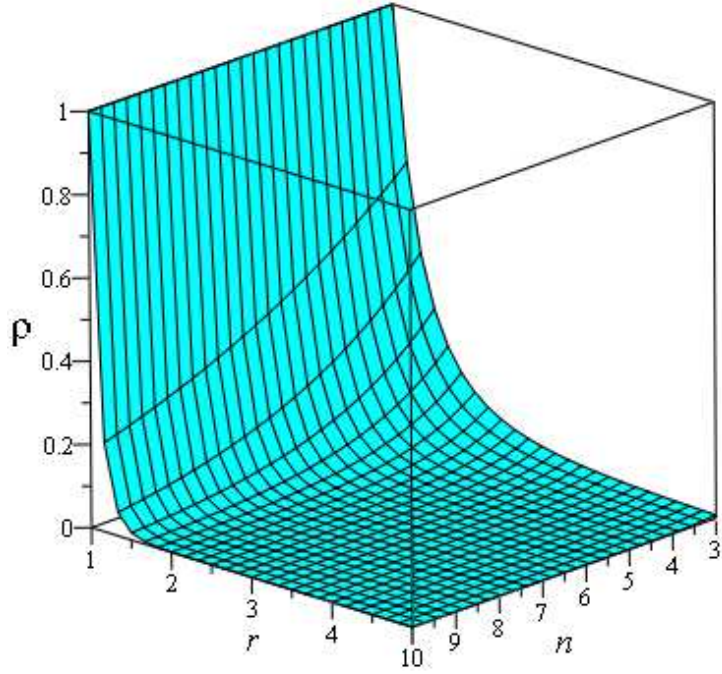}
		\caption{Variation of energy density ($\rho$) with radial co-ordinate $r$ and  parameter $n$ for the numerical values  $\lambda=10^4$, $q=1$ and $r_0=1$.}
		\label{fig3}
	\end{figure}

\begin{figure}[!htb]
	%\centering
	% \hspace*{\fill}
	\begin{minipage}{.35\textwidth}
		\centering
		\includegraphics[width=.6\linewidth]{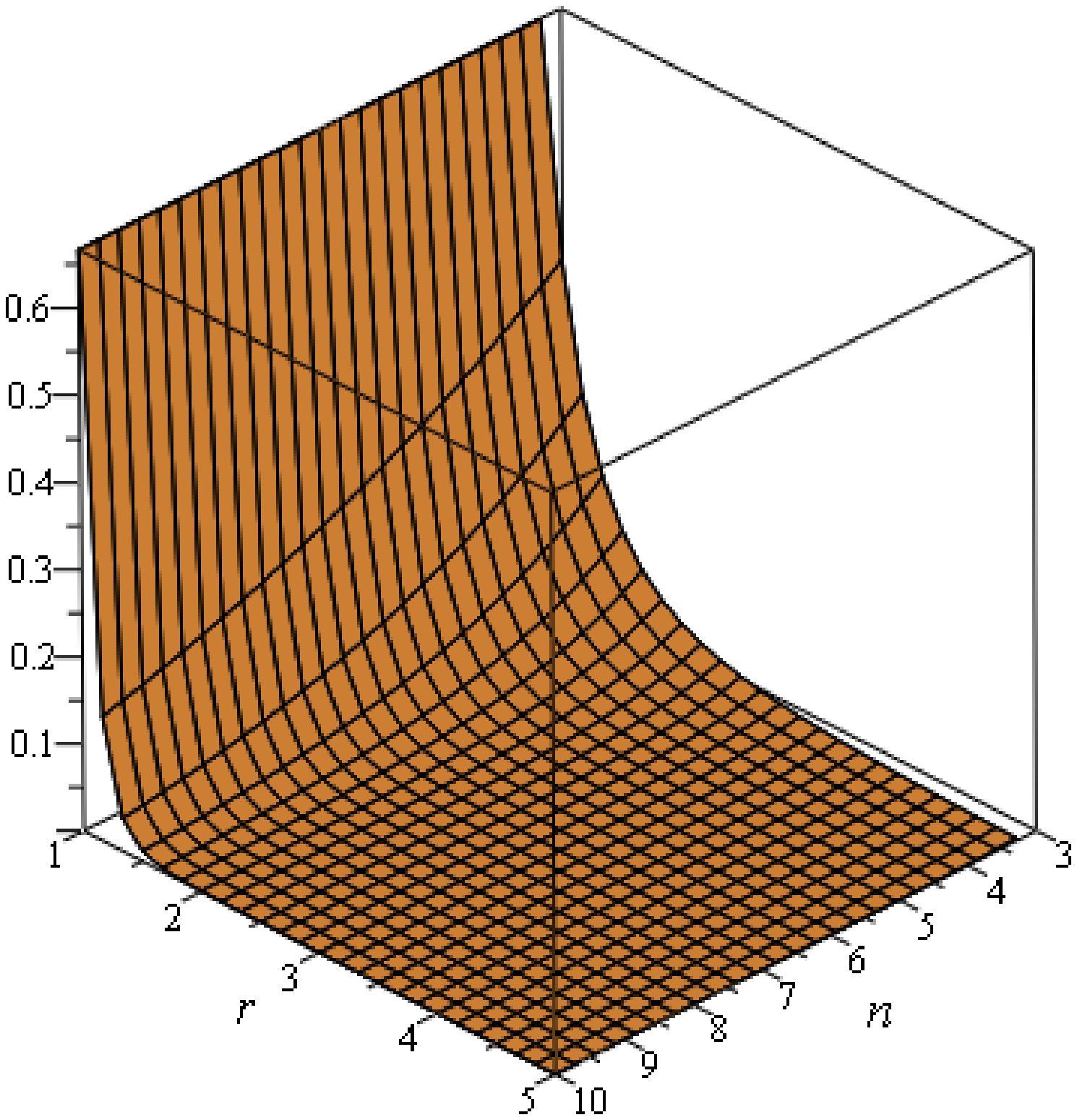}
		\centering 4(A)
	\end{minipage}%%
	\begin{minipage}{.35\textwidth}
		\centering
		\includegraphics[width=.6\linewidth]{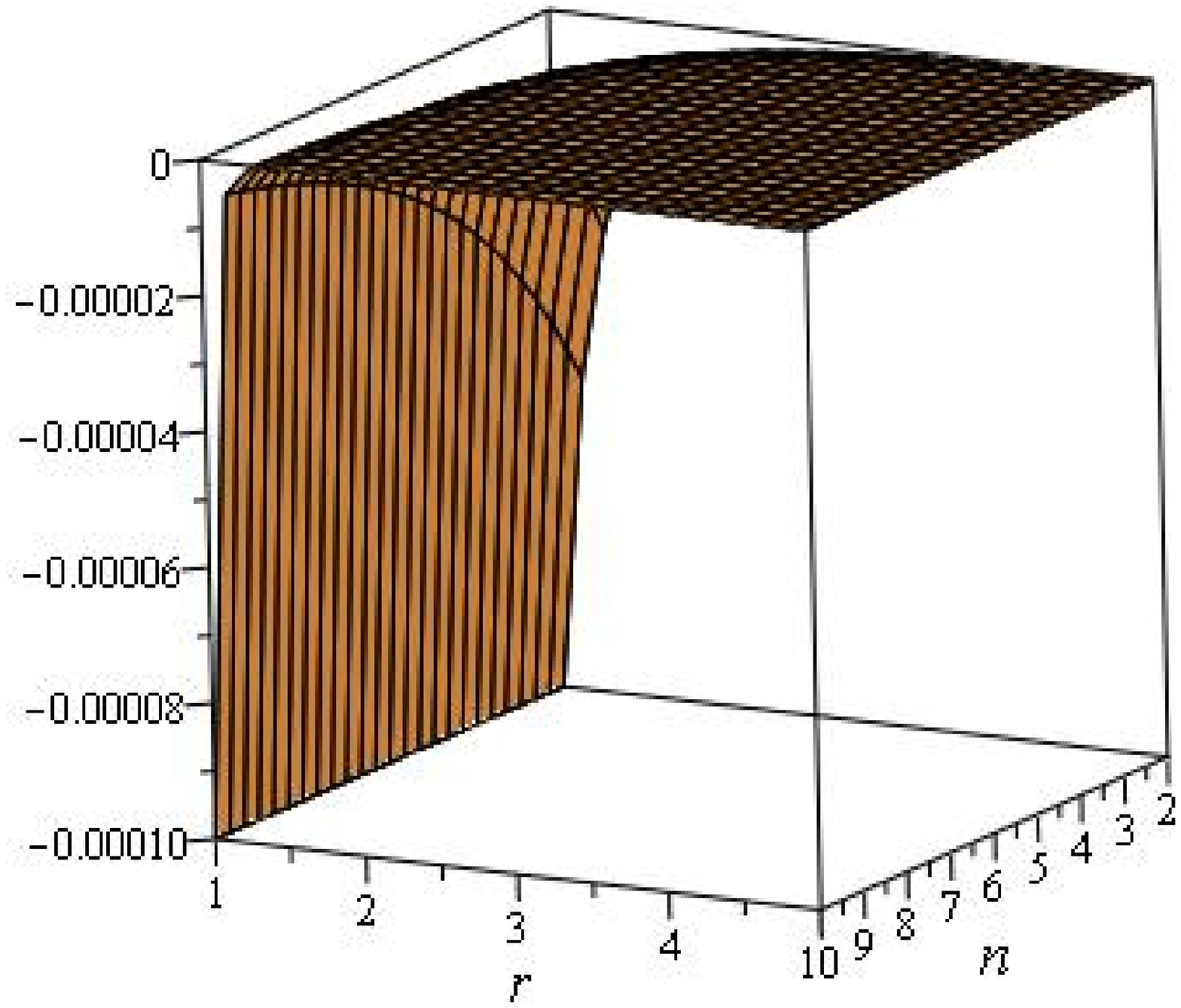}
		\centering 4(B)
	\end{minipage}%%
	\begin{minipage}{.35
			\textwidth}
		\centering
		\includegraphics[width=.6\linewidth]{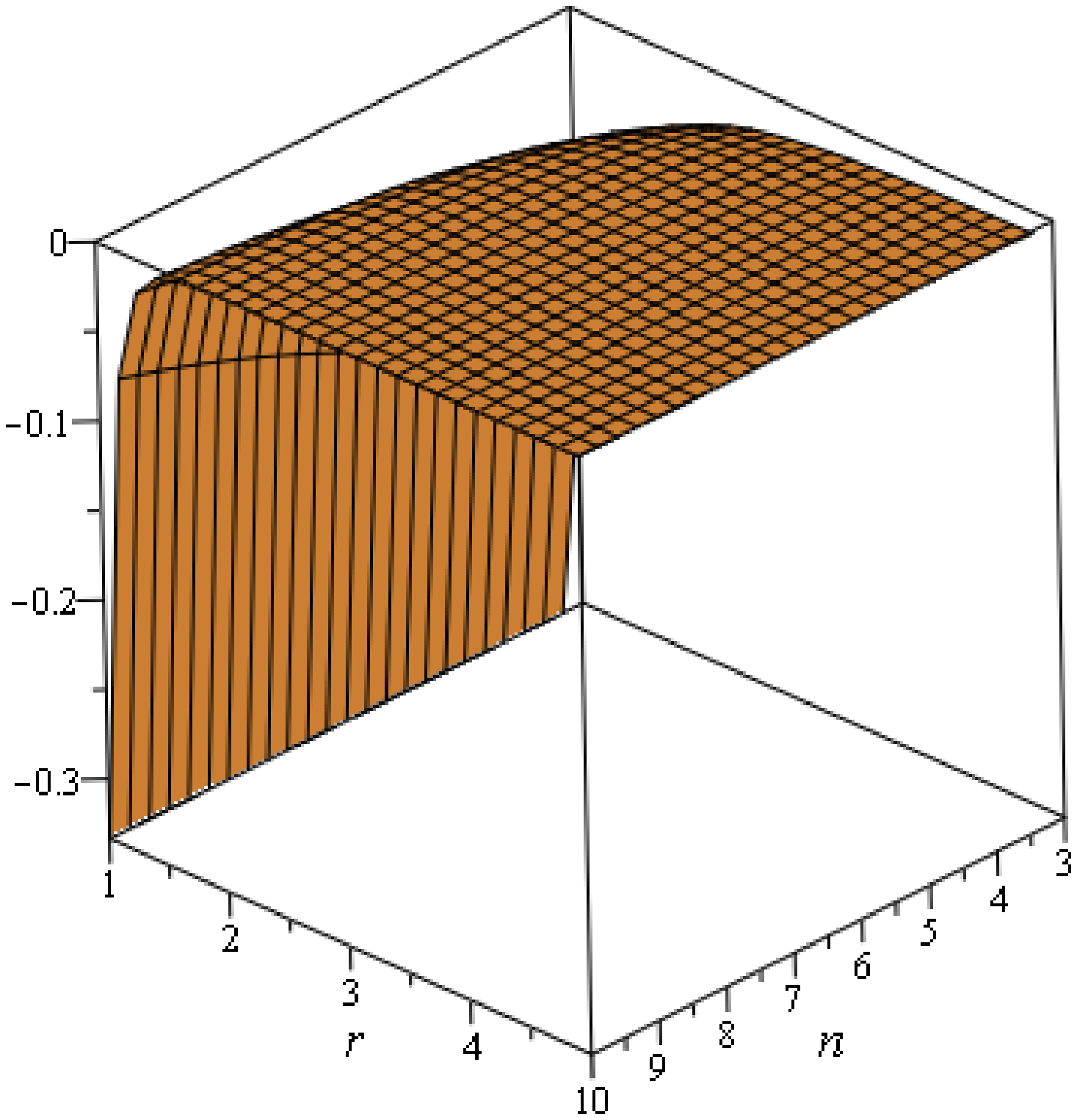}
		\centering 4(C)
	\end{minipage}%%
\caption{Variation of $\rho+p$ (4(A)), $\rho+3p$ (4(B)), $p$ (4(C)) with radial co-ordinate $r$ and  parameter $n$, have been plotted for obtained shape function(\ref{b}) with redshift function $\phi(r)=\text{constant}$ when $\lambda=10^4$, $q=1$ and $r_0=1$.}
\label{fig4}
\end{figure}

\begin{figure}
\begin{minipage}{.35\textwidth}
%	\centering
	\includegraphics[width=.6\linewidth]{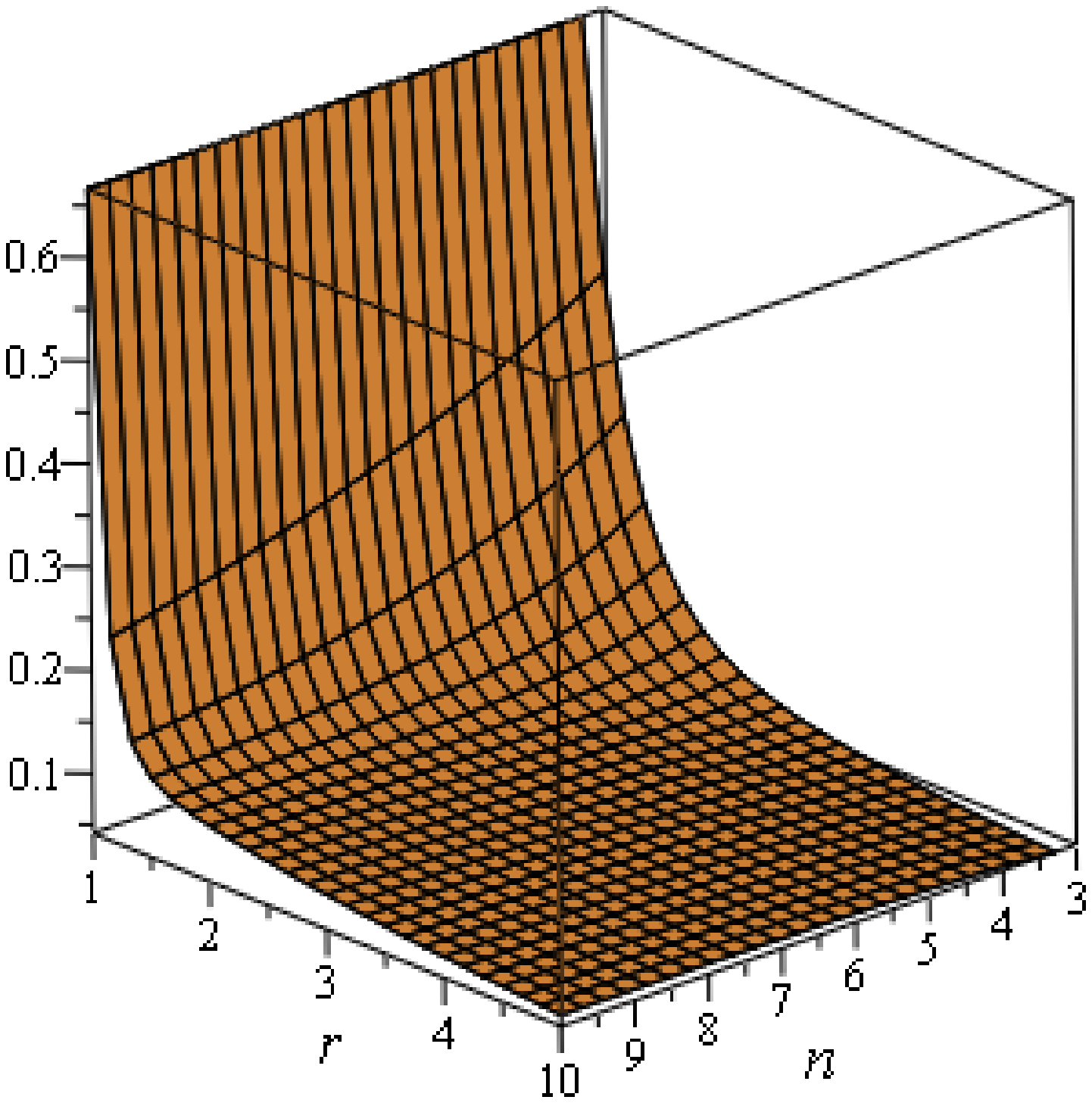}
	5(A)
\end{minipage}%%
\begin{minipage}{.35\textwidth}
	\centering
	\includegraphics[width=.6\linewidth]{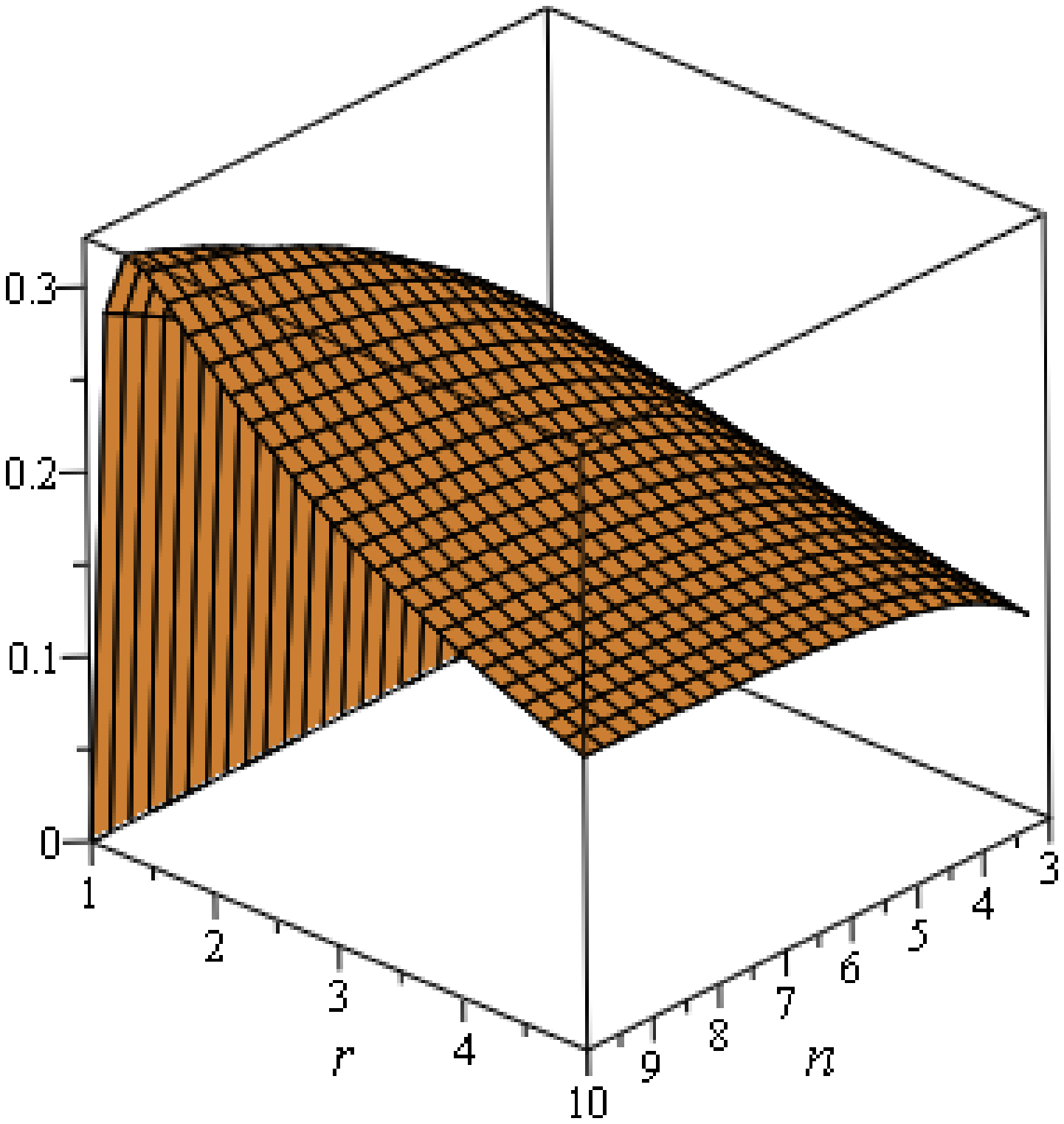}
	\centering 5(B)
\end{minipage}%%
\begin{minipage}{.35\textwidth}
	\centering
	\includegraphics[width=.6\linewidth]{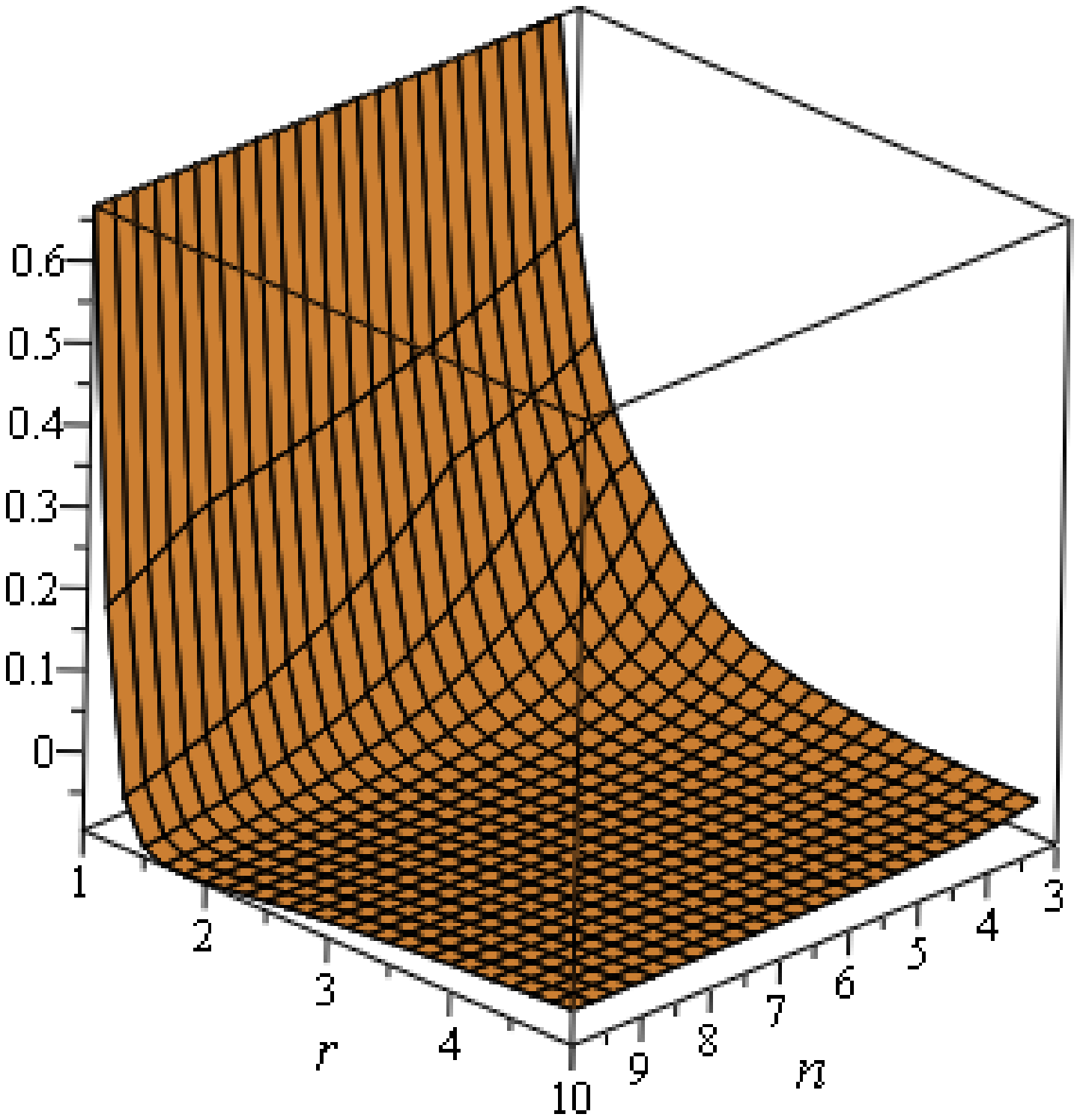}
	\centering 5(C)
\end{minipage}%%
	\caption{Variation of $\rho+p$ (5(A)), $\rho+3p$ (5(B)), $\rho-|p|$ (5(C)) with radial co-ordinate $r$ and  parameter $n$, have been plotted for obtained shape function(\ref{b}) with redshift function $\phi(r)= \beta ln\left(\frac{r}{r_0}\right)$ when $\lambda=10^4$, $\beta=0.5$, $q=1$ and $r_0=1$.}
\label{fig5}
\end{figure}

\begin{figure}[!htb]
	\centering
	\begin{minipage}{.35\textwidth}
		\centering
		\includegraphics[width=.6\linewidth]{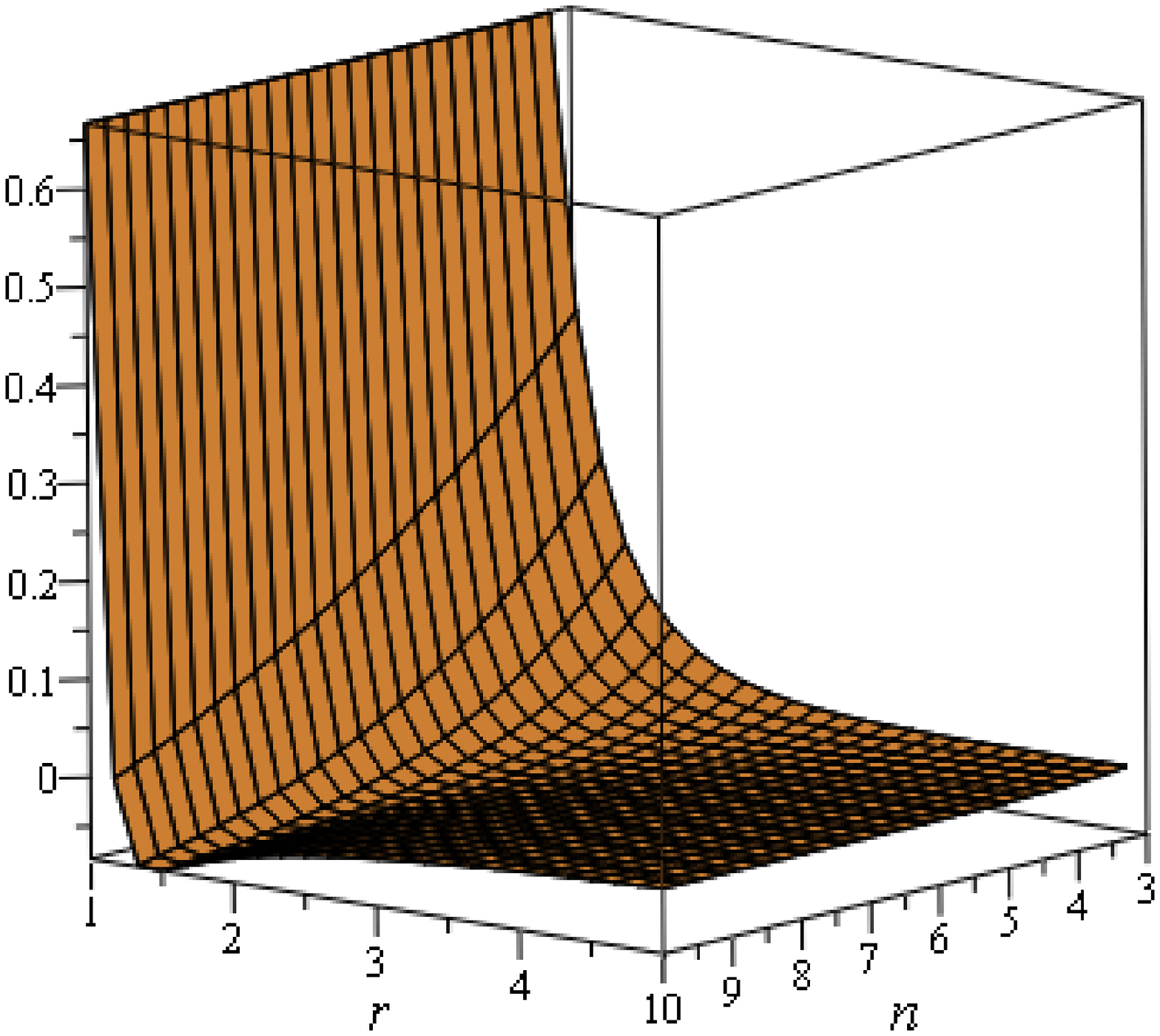}
		\centering 6(A)
	\end{minipage}%%
	\begin{minipage}{.35\textwidth}
		\centering
		\includegraphics[width=.6\linewidth]{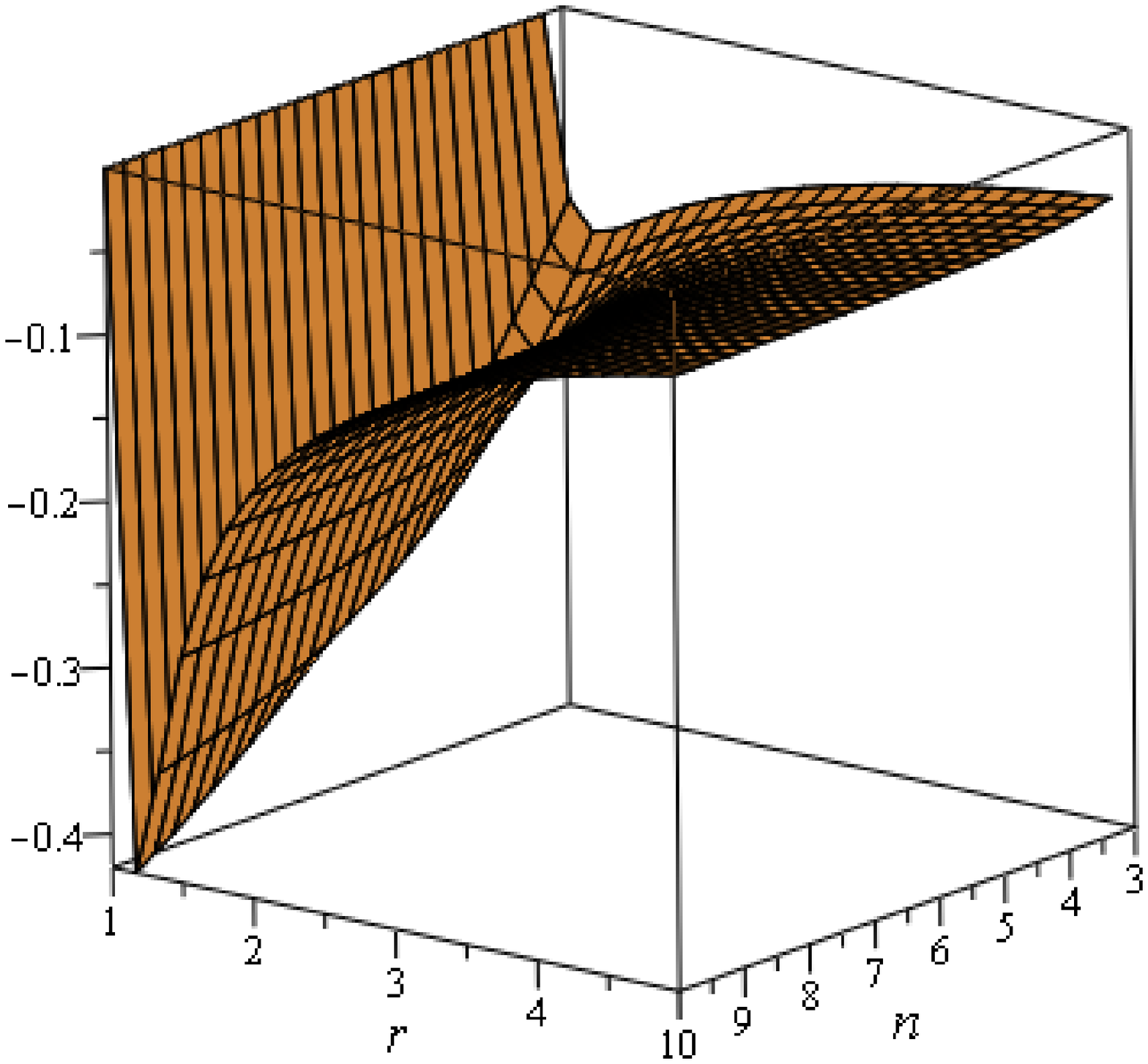}
		\centering 6(B)
	\end{minipage}%%
	\begin{minipage}{.35\textwidth}
		\centering
		\includegraphics[width=.6\linewidth]{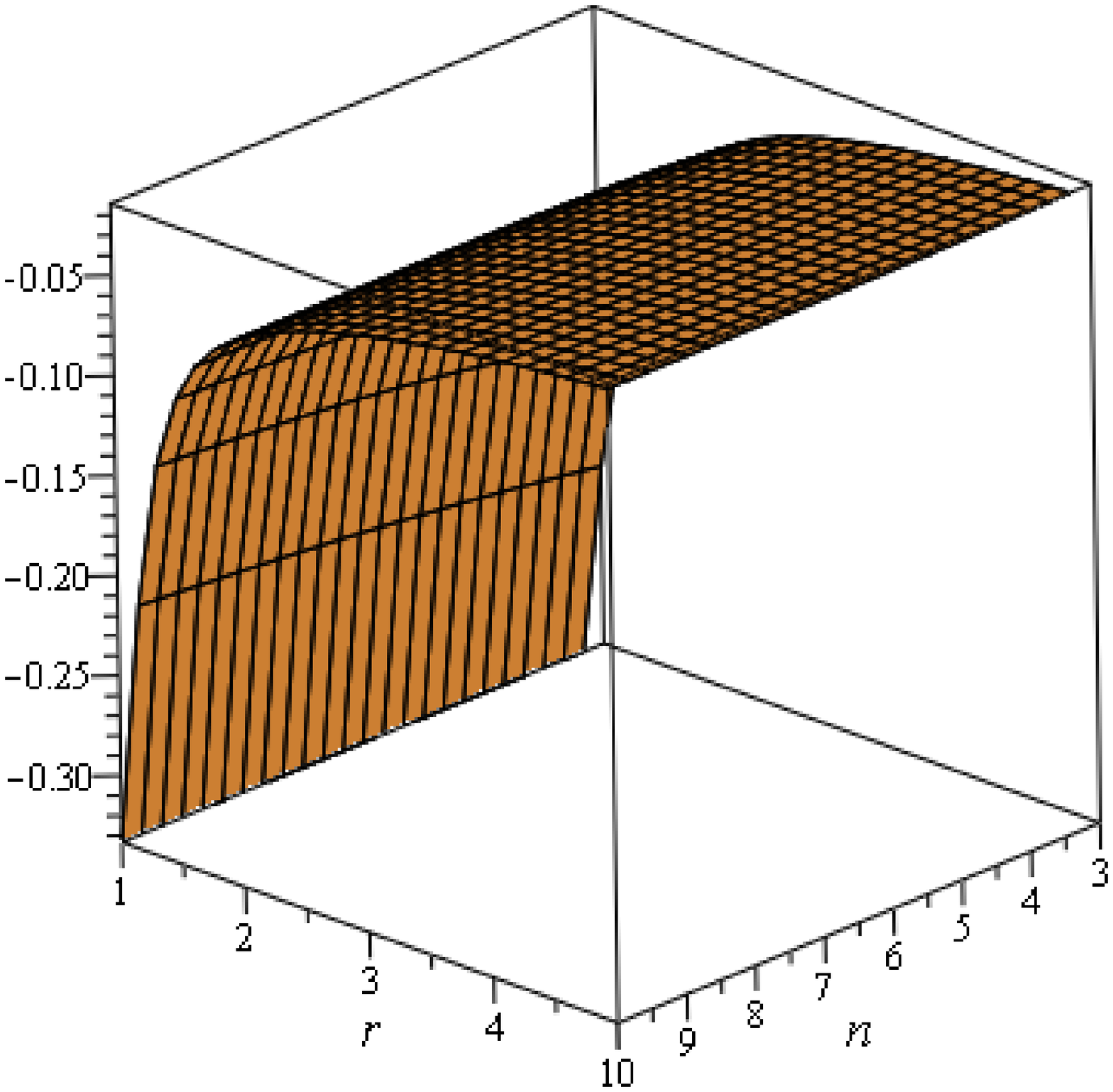}
		\centering 6(C)
	\end{minipage}%%
	\caption{Variation of $\rho+p$ (6(A)), $\rho+3p$ (6(B)), $p$ (6(C)) with radial co-ordinate $r$ and  parameter $n$, have been plotted for obtained shape function(\ref{b}) with redshift function $\phi(r)=\frac{1}{r}$ when $\lambda=10^4$, $q=1$ and $r_0=1$.}
	\label{fig6}
\end{figure}

\begin{figure}[!htb]
	\centering
	% \hspace*{\fill}
	\begin{minipage}{.35\textwidth}
		\centering
		\includegraphics[width=.6\linewidth]{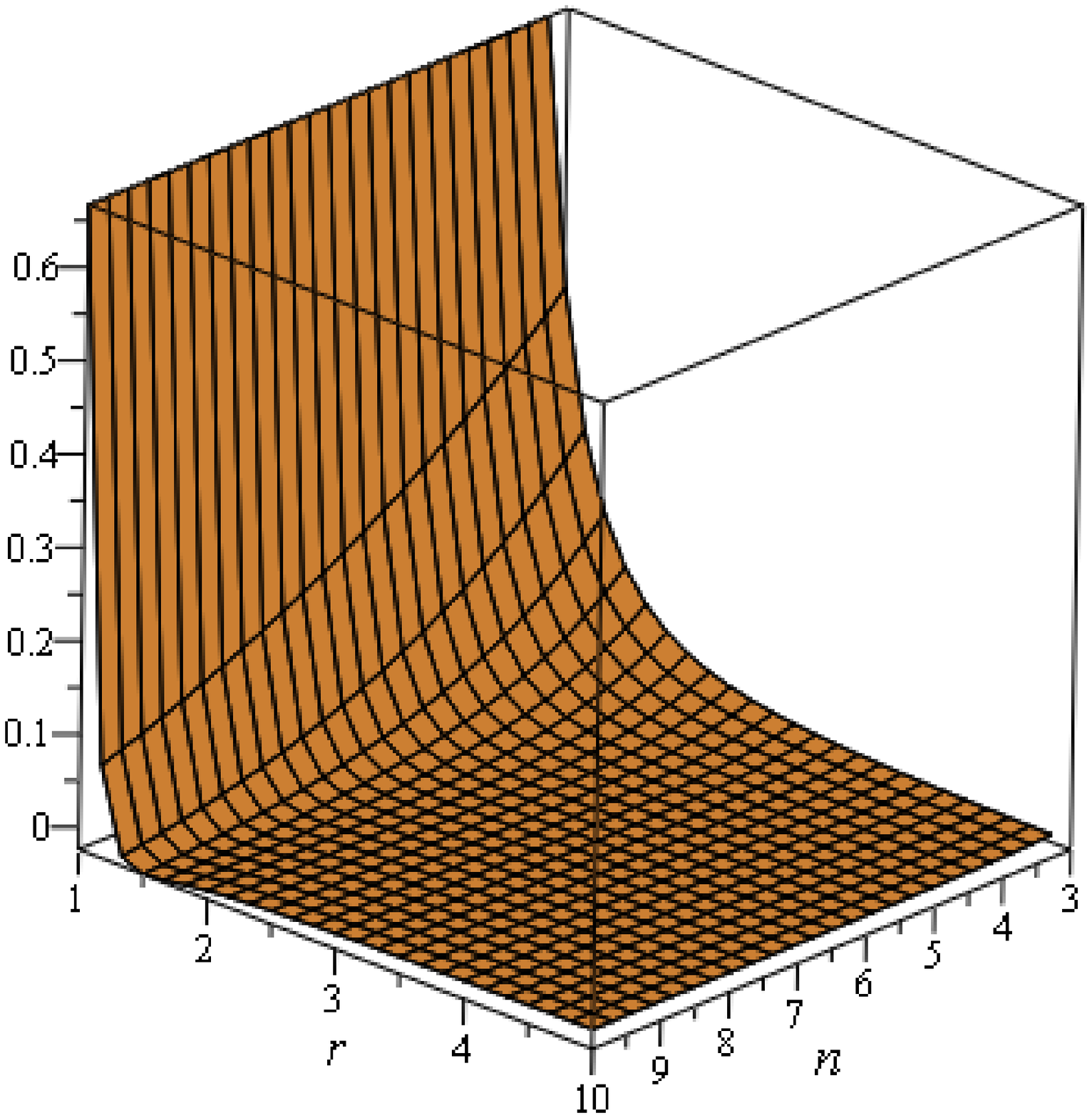}
		\centering 7(A)
	\end{minipage}%%
	\begin{minipage}{.35\textwidth}
		\centering
		\includegraphics[width=.6\linewidth]{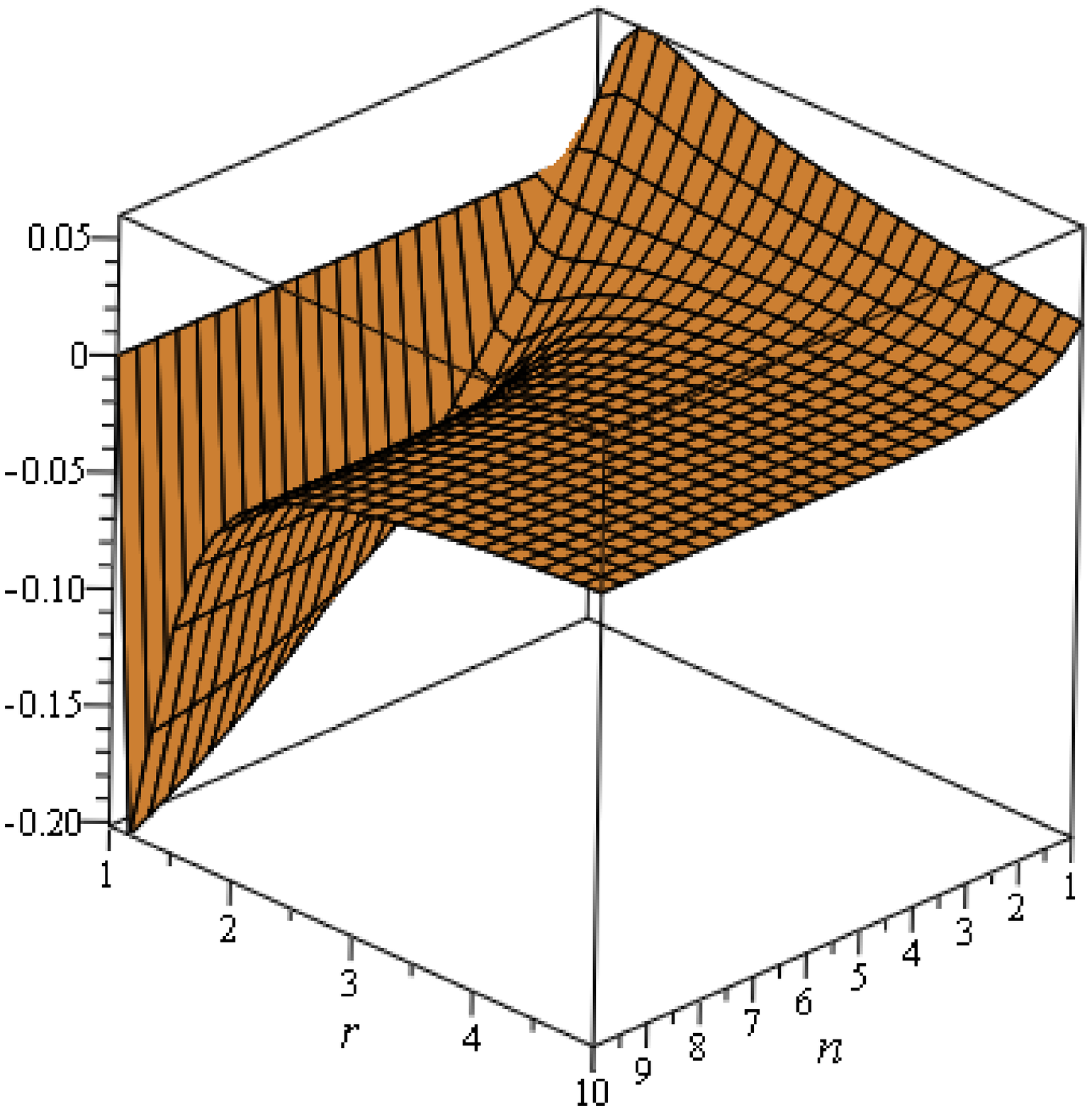}
		\centering 7(B)
	\end{minipage}%%
	\begin{minipage}{.35\textwidth}
		\centering
		\includegraphics[width=.6\linewidth]{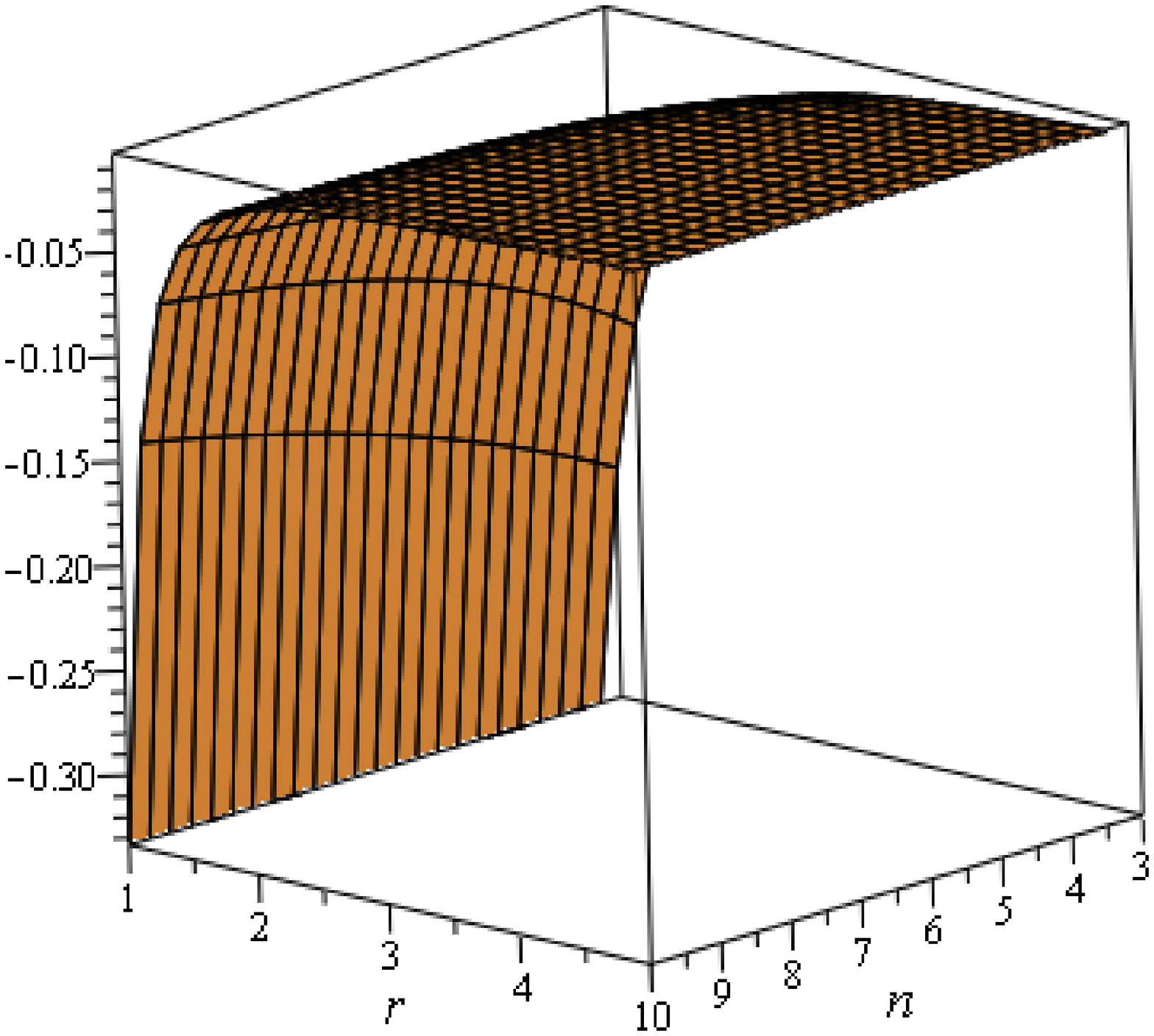}
		\centering 7(C)
	\end{minipage}%%
\caption{Variation of $\rho+p$ (7(A)), $\rho+3p$ (7(B)), $p$ (7(C)) with radial co-ordinate $r$ and  parameter $n$, have been plotted for obtained shape function(\ref{b})   with redshift function $\phi(r)=ln\frac{\sqrt{\gamma^2+r^2}}{r}$ when $\lambda=10^4$, $\gamma=1$, $q=1$ and $r_0=1$.}
\label{fig7}
\end{figure}

\begin{figure}
	\begin{minipage}{.35\textwidth}
		\centering
		\includegraphics[width=.6\linewidth]{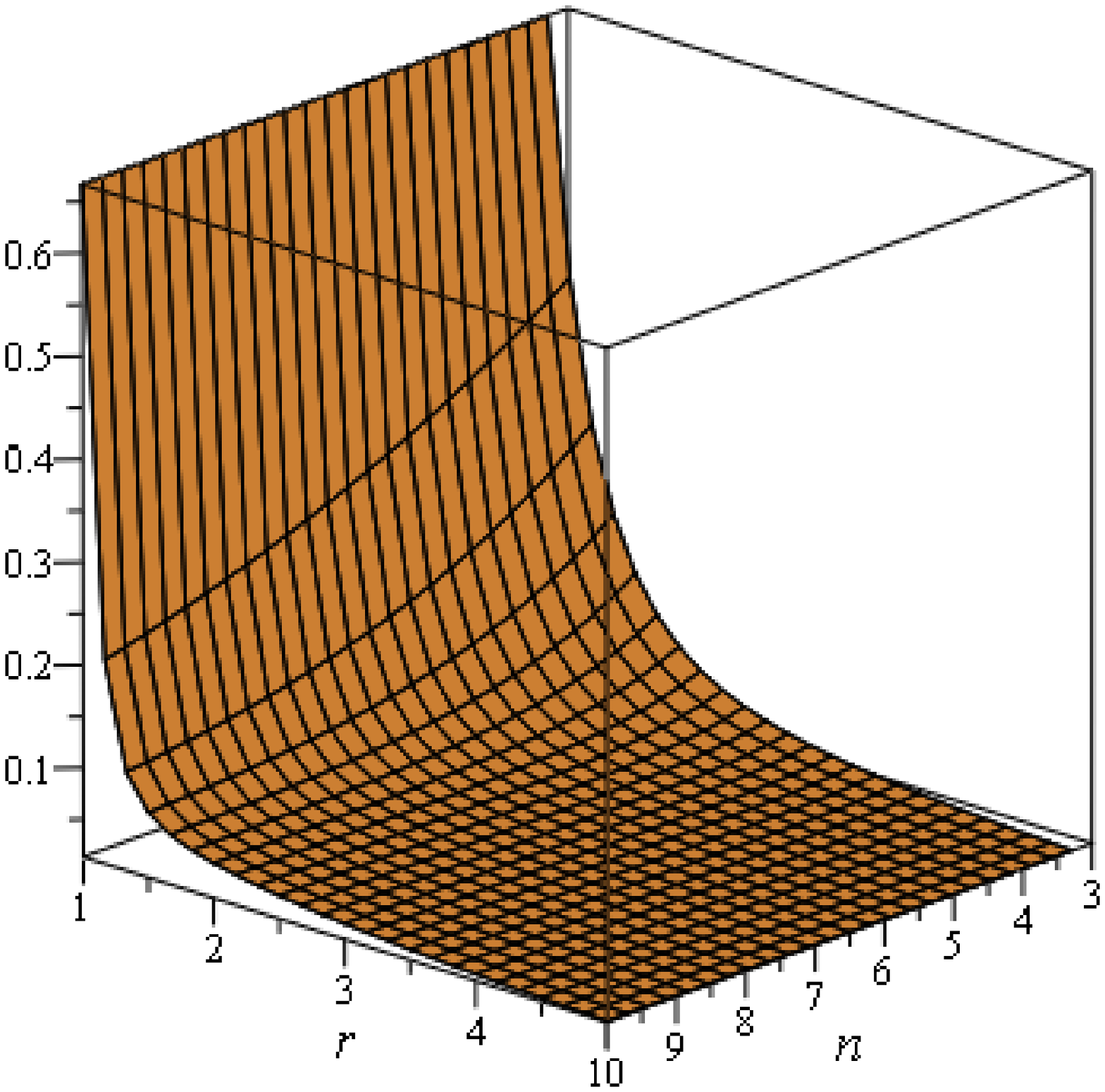}
		\centering 8(A)
	\end{minipage}%%
	\begin{minipage}{.35\textwidth}
		\centering
		\includegraphics[width=.6\linewidth]{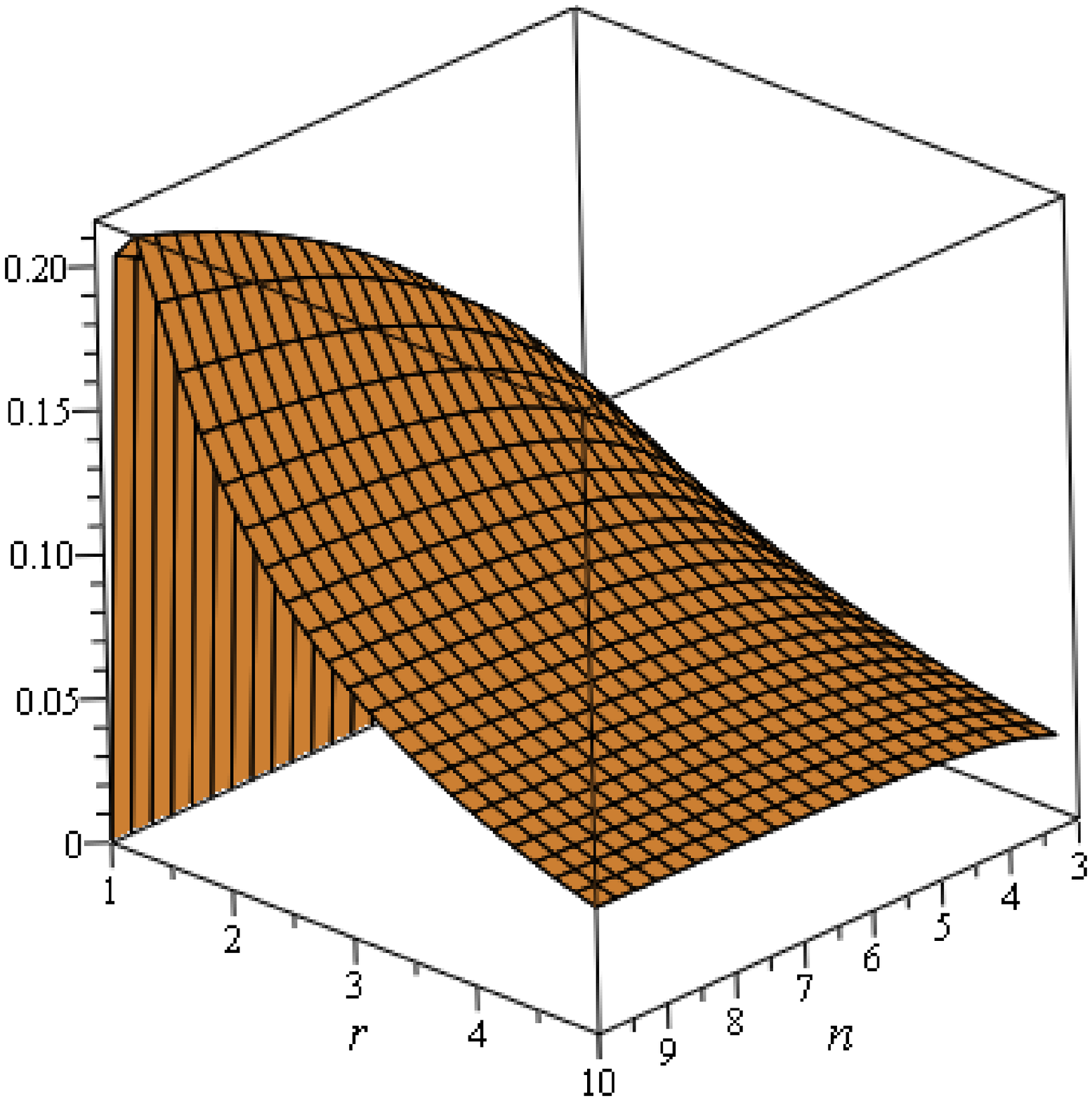}
		\centering 8(B)
	\end{minipage}%%
	\begin{minipage}{.35\textwidth}
		\centering
		\includegraphics[width=.6\linewidth]{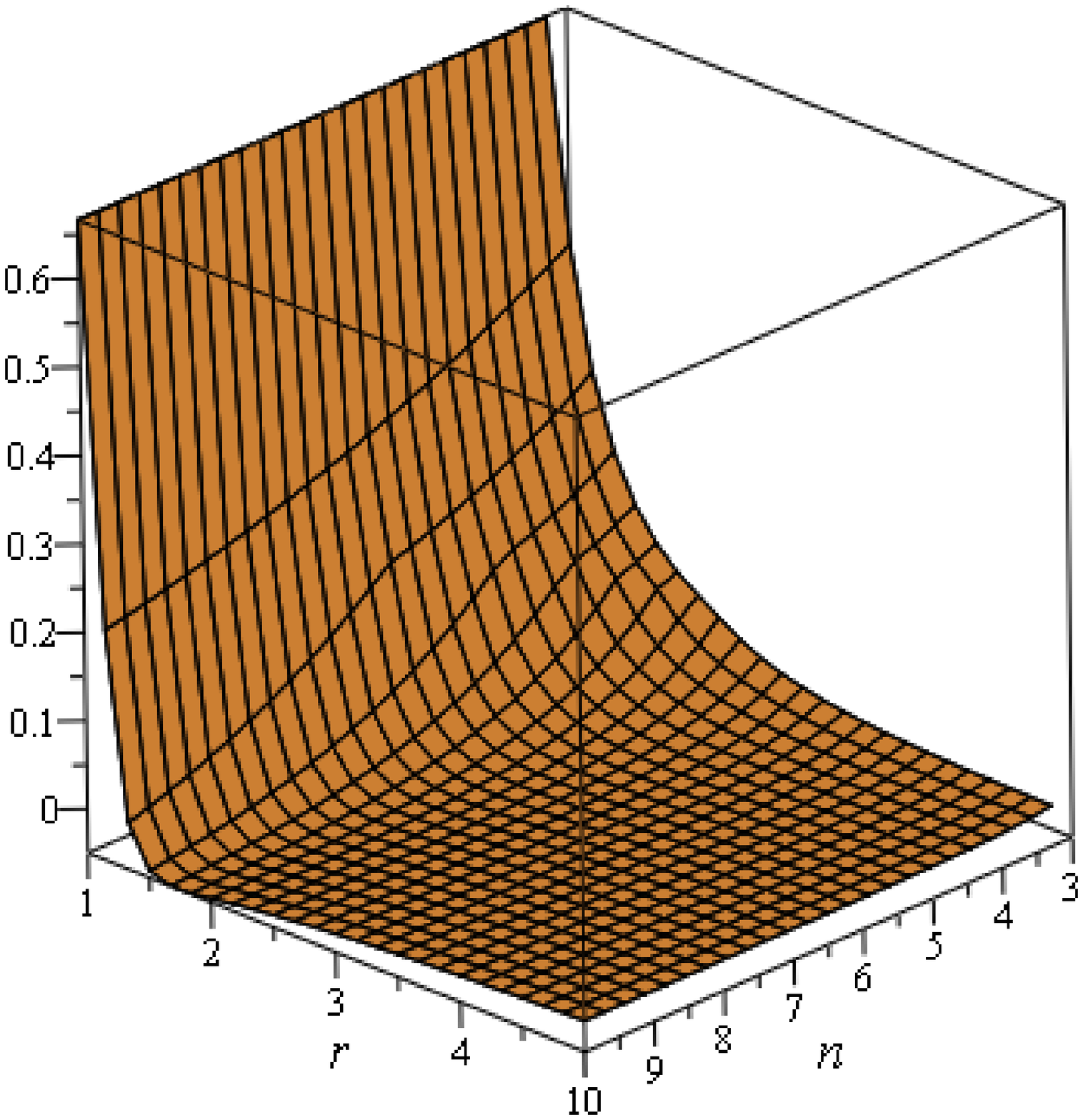}
		\centering 8(C)
	\end{minipage}%%
\caption{Variation of $\rho+p$ (8(A)), $\rho+3p$ (8(B)), $\rho-|p|$ (8(C)) with radial co-ordinate $r$ and  parameter $n$, have been plotted for obtained shape function(\ref{b})  with redshift function $\phi(r)=e^{-\frac{r_0}{r}}$ when $\lambda=10^4$, $q=1$ and $r_0=1$.}
\label{fig8}
\end{figure}

\begin{figure}
	\begin{minipage}{.35\textwidth}
		\centering
		\includegraphics[width=.6\linewidth]{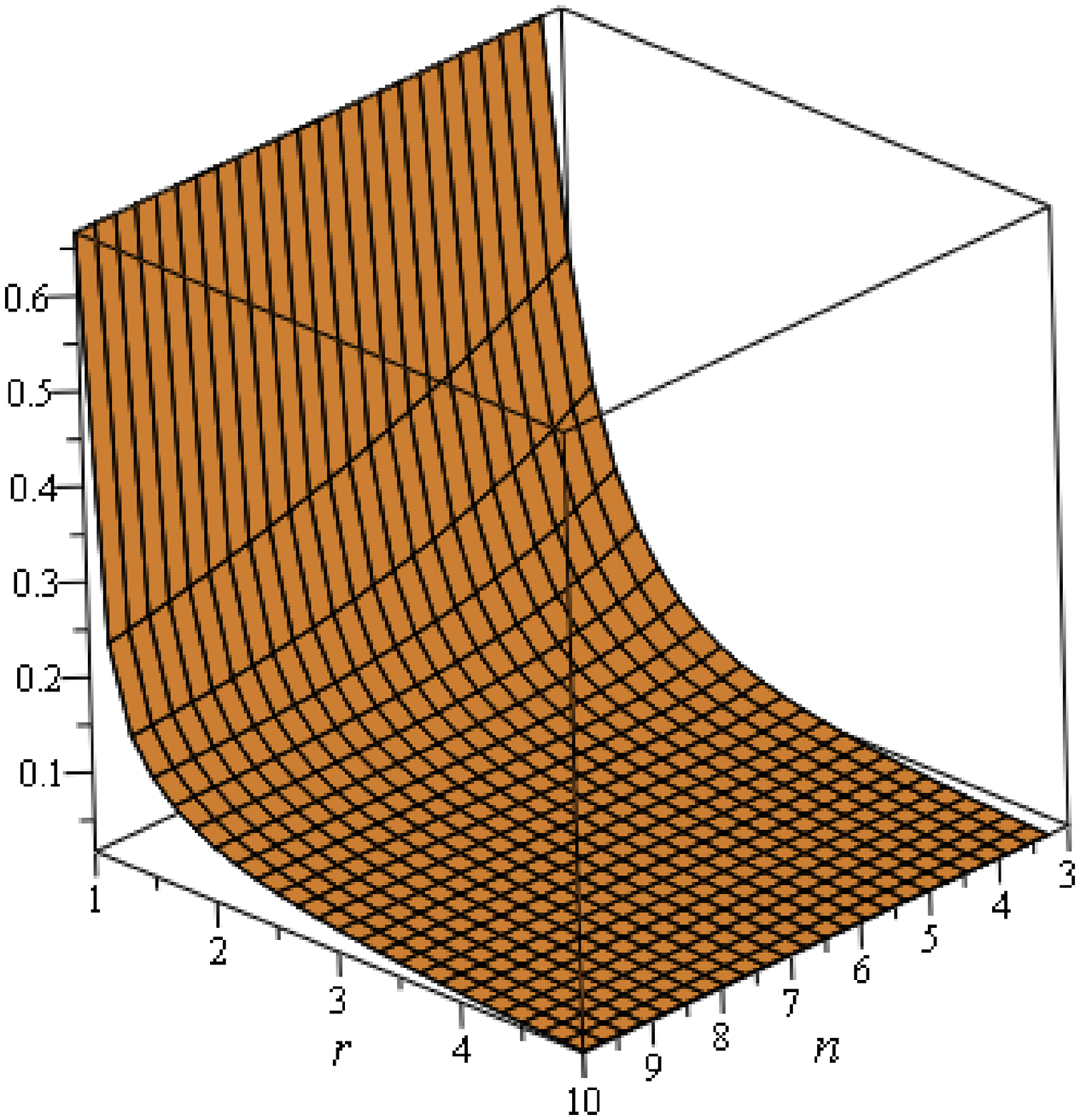}
		\centering 9(A)
	\end{minipage}%%
	\begin{minipage}{.35\textwidth}
		\centering
		\includegraphics[width=.6\linewidth]{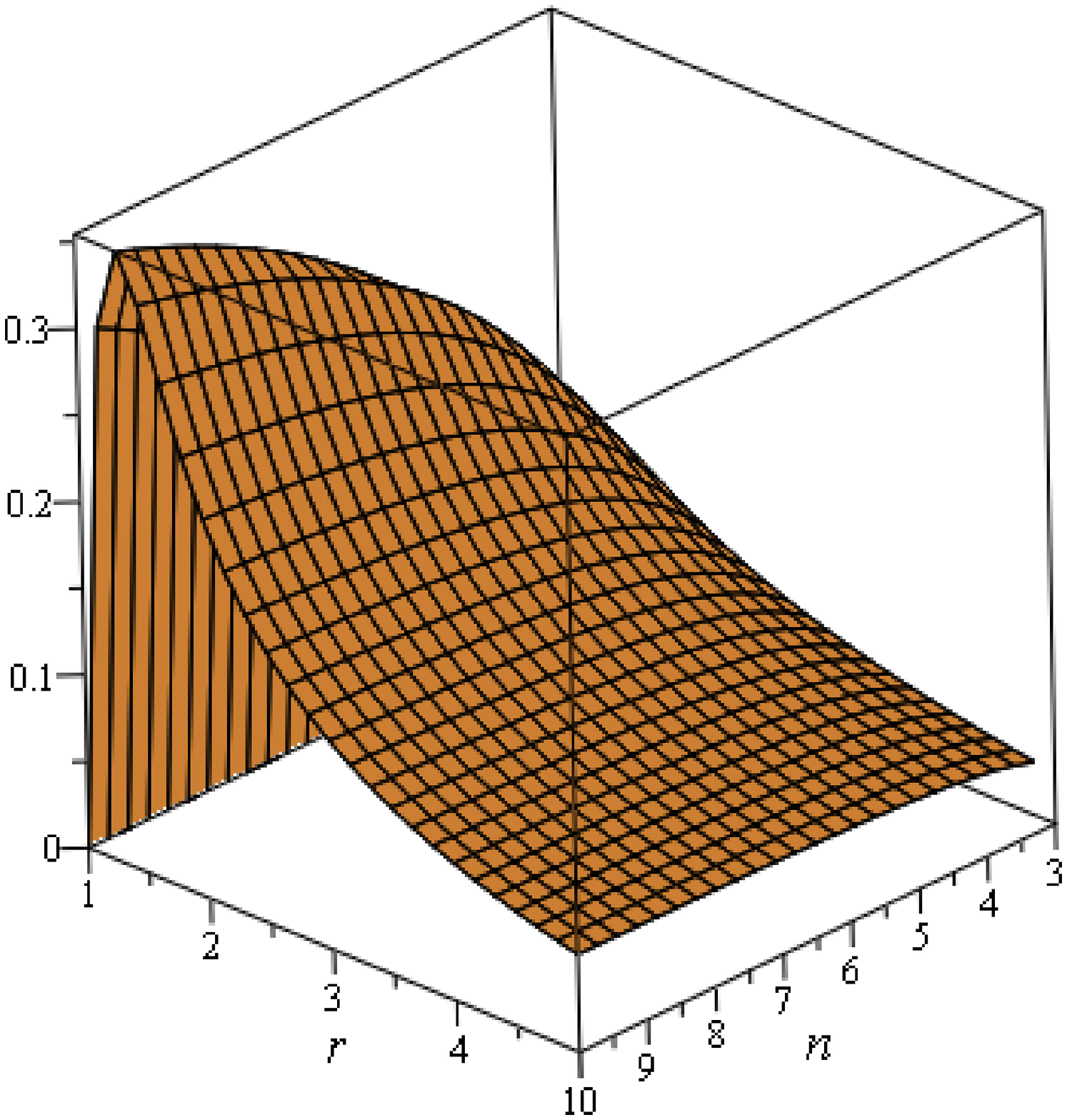}
		\centering 9(B)
	\end{minipage}%%
	\begin{minipage}{.35\textwidth}
		\centering
		\includegraphics[width=.6\linewidth]{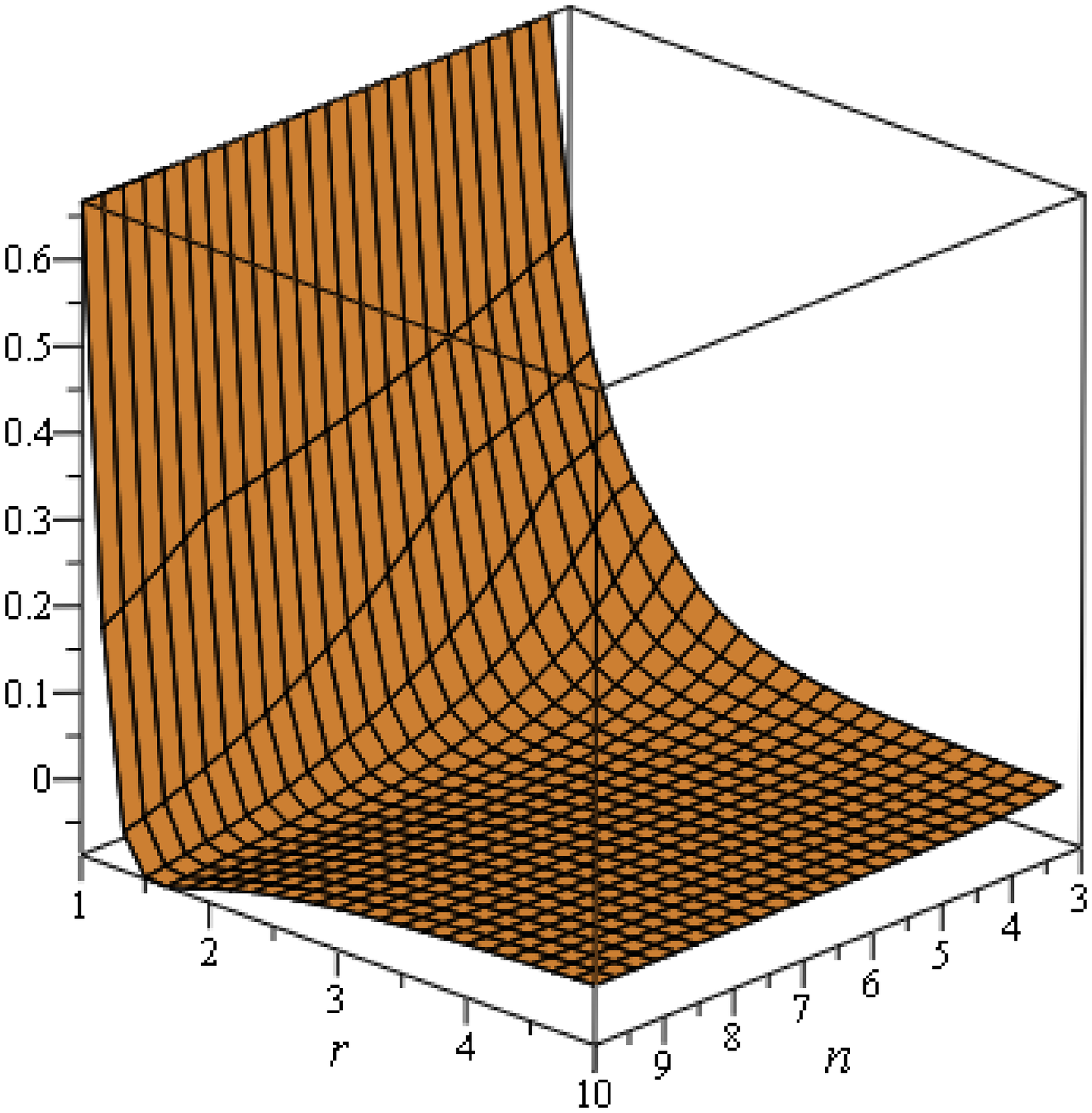}
		\centering 9(C)
	\end{minipage}%%
\caption{Variation of $\rho+p$ (9(A)), $\rho+3p$ (9(B)), $\rho-|p|$ (9(C)) with radial co-ordinate $r$ and  parameter $n$, have been plotted for obtained shape function(\ref{b})  with redshift function $\phi(r)=e^{-\frac{r_0}{r}-\frac{r_0^2}{r^2}} $  when $\lambda=10^4$, $q=1$ and $r_0=1$.}
\label{fig9}
\end{figure}

 
	Now by equation $(\ref{eq12})$ expression of $b$ $(\ref{b})$ is obtained considering $\rho=q\left({\frac{r}{r_0}}\right)^{-n}$ for $n>0$. Also equation $(\ref{eq12})$ can be written as 
	
	\begin{equation}\label{u}
	b^\prime=\frac{\rho^2r^2}{2\lambda}+\rho r^2 ~~\text{with}~ b(r_0)=r_0.
	 \end{equation}
%	  which gives a solution
%	  \begin{equation}
%	   \rho=\lambda \left( -1+\sqrt{1+\frac{2b^\prime}{\lambda r^2}}\right).
%	   \end{equation}
	   The above equation(\ref{u}) gives a unique solution for $b(r)$ in a neighbourhood of $r_0$. So if we take $\epsilon=0$, then we will get the same $\rho$ for this shape function $b(r)$.	For $\epsilon=0$, equation (\ref{eq11}) gives
	   \begin{equation}\label{r}
	   \sigma_r= - 2\sigma_t.
	   \end{equation}
	   Using equation $(\ref{r})$, equation $(\ref{eq13.2})$ and $(\ref{eq14})$ reduce to the form
	\begin{eqnarray}\label{eq27}
	A_1p-2B_1 \sigma_t&=& C_1, \\\label{eq28.1}
	A_1p+B_1 \sigma_t&=& C_2; \\
	\text{where}~A_1&=& 1+\frac{\rho}{\lambda};~B_1=\frac{-1}{8\pi}; \nonumber\\
	C_1&=& \frac{2}{r}\left(1-\frac{b}{r} \right)\phi^\prime -\frac{b}{r^3}-\frac{\rho^2}{2\lambda}; \nonumber \\
	C_2&=& \left(1-\frac{b}{r} \right)\left[\phi^{\prime \prime} +\phi^\prime (\phi^\prime +\frac{1}{r}) \right] -\frac{b^\prime r-b}{2r^2} (\phi^\prime +\frac{1}{r})-\frac{\rho^2}{2\lambda} \nonumber.
	\end{eqnarray}
	Solving the above set of equations (\ref{eq27}) and (\ref{eq28.1}) we get,
	\begin{eqnarray}
	p&=&\frac{2C_2+C_1}{3A_1}; \\
	\sigma_t&=& \frac{C_2-C_1}{3B_1} .
		\end{eqnarray}
		The shape function(\ref{b}) is obtained from the field equation (\ref{eq12}) which is independent of redshift function. Now choosing different redshift functions (which are mentioned in section-\ref{int} ), the validation of energy conditions are examined and they are shown in the figures (\ref{fig3})--(\ref{fig9}). 
	
%\begin{figure}[!]
%	\centering
%	% \hspace*{\fill}
%	\begin{minipage}{.6\textwidth}
%		\centering
%		\includegraphics[width=.6\linewidth]{pressure_redshift_1.eps}	
%	%	\centering FIG.13(A)
%	\end{minipage}%%
%	\begin{minipage}{.6\textwidth}
%		\centering
%		\includegraphics[width=.6\linewidth]{comb_redshift_1.eps}
%	%	\centering FIG.13(B)	
%	\end{minipage}
%
%	\caption{Behavior of $\rho+3p$ (left diagram) and $\rho$, $p$, $\rho+p$, and $\rho-|p|$ diagrams (right diagram) have been plotted for obtained shape function(\ref{b}) with redshift function $\phi(r)=\text{constant} $, against $r$  when $ n=3.6$, $l=10^4$ and $r_0=1.5$ .}
%	\label{fig2}
%\end{figure}

%\begin{figure}[!]
%	\centering
%	\includegraphics[width=.4\linewidth]{comb_redshift_2.eps}
%	\caption{Behavior of $\rho$, $\rho+p$, $\rho+3p$ and $\rho-|p|$ diagrams have been plotted for obtained shape function(\ref{b}) with redshift function $\phi(r)= \gamma ln\left(\frac{r}{r_0}\right)$, against $r$  when $ n=3.6$, $l=10^4$, $\gamma=05$ and $r_0=1.5$ .}
%	\label{fig3}
%\end{figure}
%\begin{figure}[h]
%	\centering
%	% \hspace*{\fill}
%	\begin{minipage}{.6\textwidth}
%		\centering
%		\includegraphics[width=.6\linewidth]{pressure_redshift_3.eps}	
%		%	\centering FIG.13(A)
%	\end{minipage}%%
%	\begin{minipage}{.6\textwidth}
%		\centering
%		\includegraphics[width=.6\linewidth]{comb_redshift_3.eps}
%		%	\centering FIG.13(B)	
%	\end{minipage}
%	
%	
%	\caption{Behavior of isotropic pressure ($p$) (left diagram) and $\rho$, $\rho+p$, $\rho+3p$ and $\rho-|p|$ diagrams (right diagram) have been plotted for obtained shape function(\ref{b}) with redshift function $\phi(r)=\frac{1}{r} $, against $r$  when $ n=3.6$, $l=10^4$ and $r_0=1.5$ .}
%	\label{fig4}
%\end{figure}

%\begin{figure}[h]
%	\centering
%	% \hspace*{\fill}
%	\begin{minipage}{.6\textwidth}
%		\centering
%		\includegraphics[width=.6\linewidth]{pressure_redshift_4.eps}	
%		%	\centering FIG.13(A)
%	\end{minipage}%%
%	\begin{minipage}{.6\textwidth}
%		\centering
%		\includegraphics[width=.6\linewidth]{comb_redshift_4.eps}
%		%	\centering FIG.13(B)	
%	\end{minipage}
%	
%	
%	\caption{Behavior of isotropic pressure ($p$) (left diagram) and $\rho$, $\rho+p$, $\rho+3p$ and $\rho-|p|$ diagrams (right diagram) have been plotted for obtained shape function(\ref{b}) with redshift function $\phi(r)= ln\frac{\sqrt{\gamma^2+r^2}}{r}$, against $r$  when $ n=3.6$, $l=10^4$, $\gamma=1$ and $r_0=1.5$ .}
%	\label{fig5}
%\end{figure}

%\begin{figure}[h]
%	\centering
%	\includegraphics[width=.4\linewidth]{comb_redshift_5.eps}
%	\caption{Behavior of $\rho$, $\rho+p$, $\rho+3p$ and $\rho-|p|$ diagrams have been plotted for obtained shape function(\ref{b}) with redshift function $\phi(r)=e^{-\frac{r_0}{r}}
%		 $ against $r$  when $ n=3.6$, $l=10^4$ and $r_0=1.5$ .}
%	 \label{fig6}
%\end{figure}
%\begin{figure}[h]
%	\centering
%	\includegraphics[width=.4\linewidth]{comb_redshift_6.eps}
%	\caption{Behavior of $\rho$, $\rho+p$, $\rho+3p$ and $\rho-|p|$ diagrams have been plotted for obtained shape function(\ref{b}) with redshift function $\phi(r)=\phi(r)=e^{-\frac{r_0}{r}-\frac{r_0^2}{r^2}} $ against $r$  when $ n=3.6$, $l=10^4$ and $r_0=1.5$ .}
%	\label{fig7}
%\end{figure}
\section{Wormhole solutions from field equations for anisotropic fluid}
\label{wsa}
We assume the relation $p_r=\alpha p_t$ ($\alpha\neq1$) and consider a equation of state $p_t=\omega\rho$ {\it i.e.,} $p_r=\alpha\omega\rho$. Now  we will find the wormhole solution from the above field equations (\ref{eq7})-(\ref{eq9}). Using above equation of state in (\ref{eq10}) we obtain ,
\begin{equation}
R=-k^2\biggl[(-1+\alpha\omega+2\omega)\rho+\frac{\rho^2}{\lambda}(1+\alpha^2\omega^2+\omega^2+\alpha\omega+2\omega-2\alpha\omega^2)\biggr]
\end{equation}
{\it i.e.},
\begin{equation}\label{eq13}
\frac{1+\alpha^2\omega^2+\omega^2+\alpha\omega+2\omega-2\alpha\omega^2}{\lambda}\rho^2+(-1+\alpha\omega+2\omega)\rho+\frac{1}{8\pi}R=0.
\end{equation}
After solving the equation(\ref{eq13}), we get
	\begin{equation}\label{eq19}
\rho(R) =\frac{-B\pm\sqrt{B^2-4AC} }{2A},~~~{\text{where}} 
\begin{cases}
A=\frac{1+\alpha^2\omega^2+\omega^2+\alpha\omega+2\omega-2\alpha\omega^2}{\lambda} \\
B=(-1+\alpha\omega+2\omega)\\
C=\frac{1}{8\pi}R
\end{cases}.
\end{equation}
Consequently we have $p_r$ and $p_t$(in terms of $\rho$). Thus from the value of $R$, the expression of $\rho$, $p_r$ and $p_t$ can be computed. %The above $\rho$ ( in \ref{eq19}) will be treated as model I. 
 In this section wormhole solutions will be investigated with five different shape functions considering the redshift function $\phi(r)=\beta ln(\frac{r}{r_0})$, where $\beta$ is an arbitrary constant: 
\subsubsection{shape function $b(r)=\frac{r_0^n}{r^{n-1}}$, for some $n>0$}\label{s1}
For shape function (1), from the equation (\ref{eq11.1}) we get the result,
\begin{equation}\label{eq22}
R=-2\Big(1-\frac{r_0^n}{r^n}\Big)\Biggl[-\frac{\beta^2}{r^2}+\frac{\beta}{r^2}+\frac{(n-1)r_0^n}{r^{n+2}\bigl(1-\frac{r_0^n}{r^n}\bigr)}-\frac{\beta}{2}\frac{\bigl\{\frac{(1-n)r_0^n}{r^{n-1}}+\frac{3r_0^n}{r^{n-1}}-4r\bigr\}}{r^3\bigl(1-\frac{r_0^n}{r^n}\bigr)}\Biggr].
\end{equation}
Therefore from (\ref{eq19}) we obtain the expression of $\rho$ for this model, 
\begin{eqnarray}\label{eq23}
\rho&=&\frac{\lambda}{2(1+\alpha^2\omega^2+\omega^2+\alpha\omega+2\omega-2\alpha\omega^2)}\Bigg[(1-\alpha\omega-2\omega)
+\Biggl\{(-1+\alpha\omega+2\omega)^2\\\nonumber
&~&+\frac{(1+\alpha^2\omega^2+\omega^2+\alpha\omega+2\omega-2\alpha\omega^2)\Big(1-\frac{r_0^n}{r^n}\Big)\Bigg\{-\frac{\beta^2}{r^2}+\frac{\beta}{r^2}+\frac{(n-1)r_0^n}{r^{n+2}\bigl(1-\frac{r_0^n}{r^n}\bigr)}-\frac{\beta}{2}\frac{\biggl\{\frac{(1-n)r_0^n}{r^{n-1}}+\frac{3r_0^n}{r^{n-1}}-4r\biggr\}}{r^3\bigl(1-\frac{r_0^n}{r^n}\bigr)}\Biggr\}}{\pi\lambda}\Biggl\}^{1/2}\Bigg].\\\nonumber
%p_t&=&\omega\rho,\\
%p_r&=&\alpha\omega\rho.
\end{eqnarray}
\begin{figure}[h]
	\centering
	% \hspace*{\fill}
	\begin{minipage}{.55\textwidth}
		\centering
		\includegraphics[width=.6\linewidth]{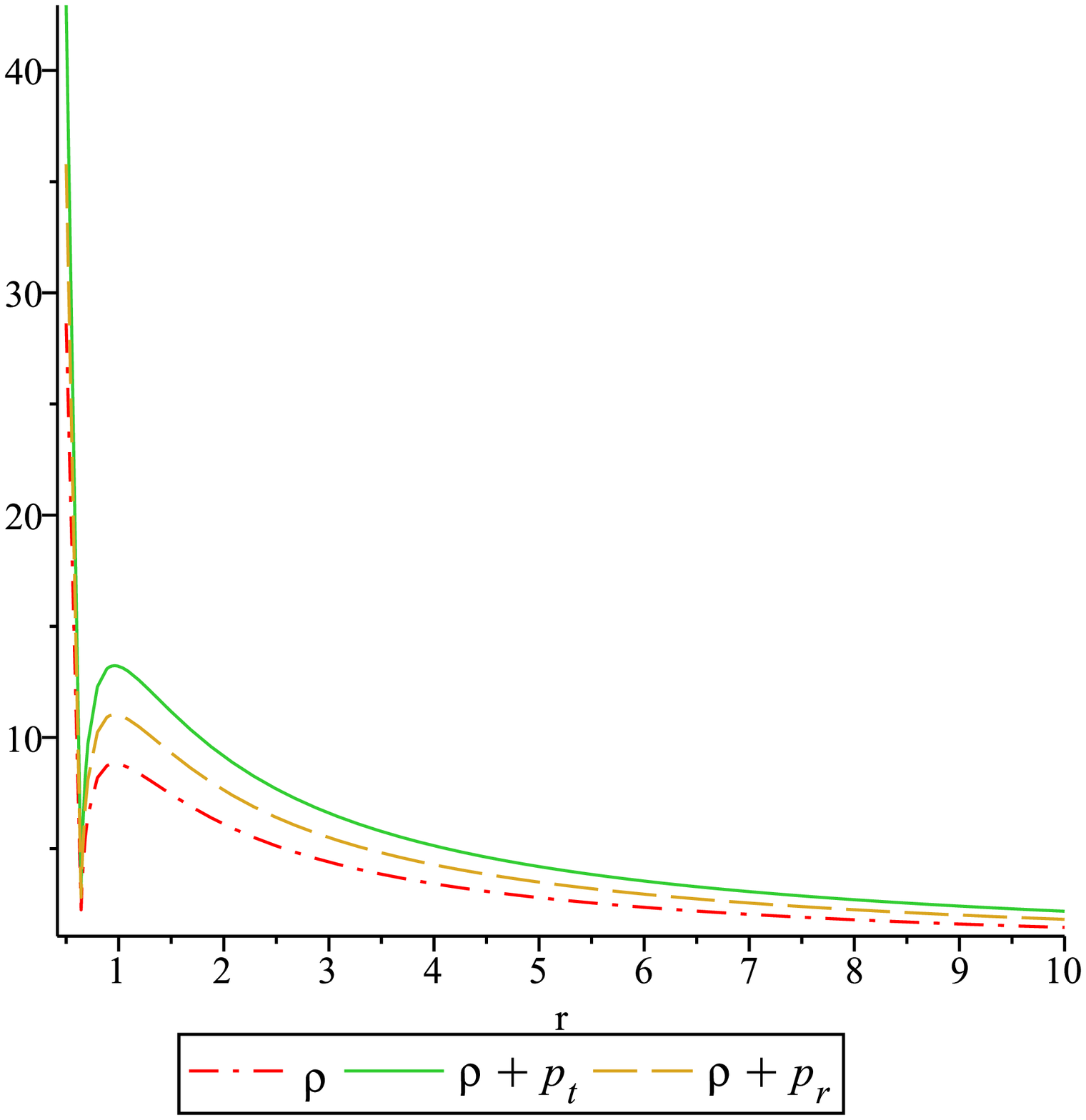}
		\centering FIG.10(A)
	\end{minipage}%%
	\begin{minipage}{.55\textwidth}
		\centering
		\includegraphics[width=.6\linewidth]{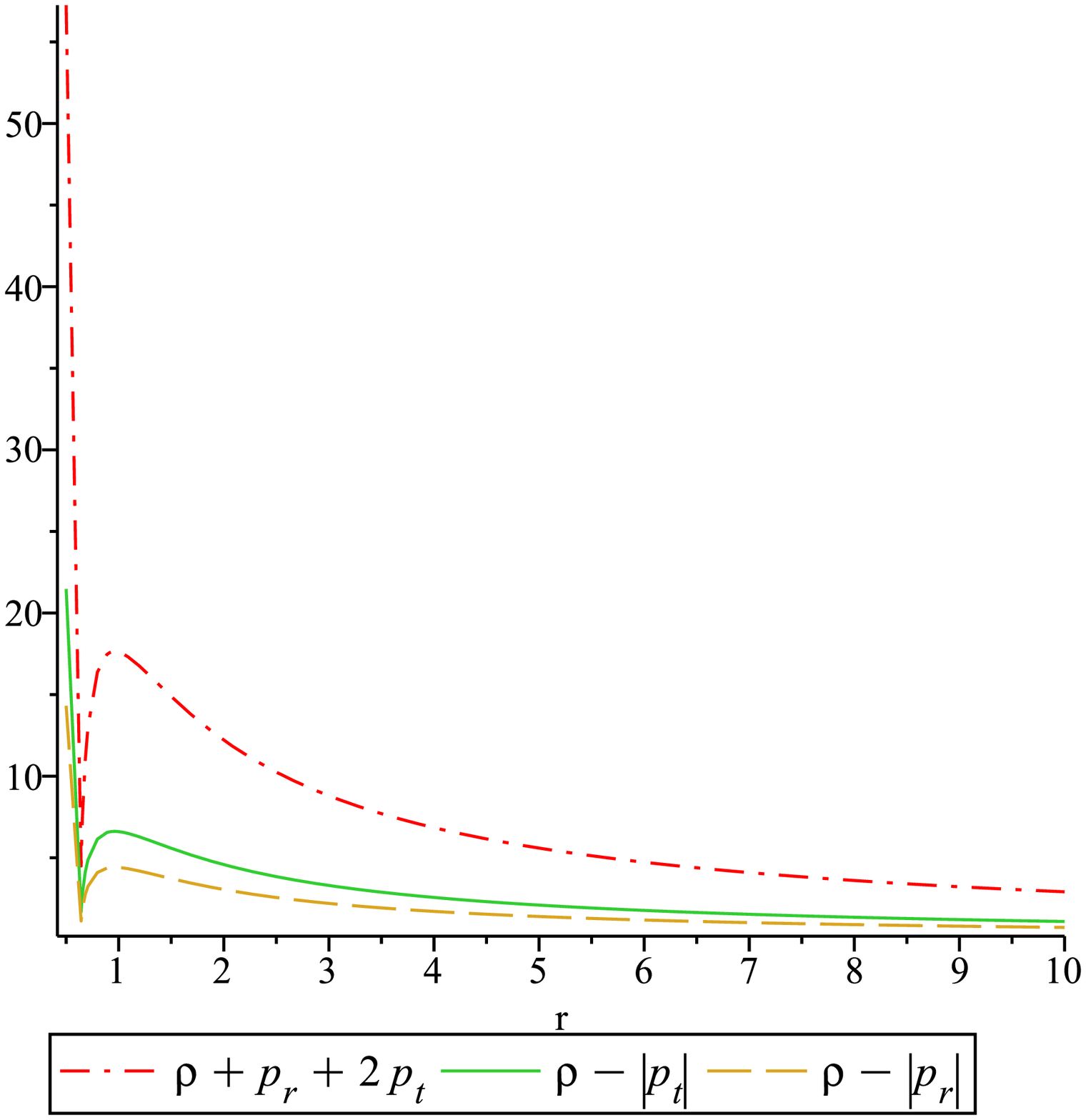}
		\centering FIG.10(B)
	\end{minipage}
	
	\caption{Behavior of $\rho$, $\rho+p_t$ and $\rho+p_r$ (FIG.10(A)) and $\rho+p_r+2p_t$, $\rho-|p_t|$ and $\rho-|p_r|$ diagrams (FIG.10(B)) have been plotted for shape function (1
		) with non-zero tidal force against $r$ when $\alpha=2$, $\omega=0.25$, $\lambda=10^3$, $n=0.9$, $\beta=-3$ and $r_0=0.5$.}
	\label{fig10}
\end{figure}
\subsubsection{shape function $b(r)=\frac{r}{1+ln(1+r-r_0)}$, $0<r_0<1$}
For shape function (2), from the equation (\ref{eq11.1}) we get the result,
\begin{eqnarray}
\label{eq25}\nonumber
R&=&\frac{2ln(1+r-r_0)}{1+ln(1+r-r_0)}\Bigg[\frac{\beta}{r^2}-\frac{\beta^2}{r^2}+\frac{(1-r_0)+(1+r-r_0)ln(1+r-r_0)}{r(1+r-r_0)(1+ln(1+r-r_0))ln(1+r-r_0)}\\
&+&\frac{\beta(1+ln(1+r-r_0))}{2r^2ln(1+r-r_0)}\Biggl\{\frac{(1-r_0)+(1+r-r_0)ln(1+r-r_0)}{(1+ln(1+r-r_0))^2(1+r-r_0)}+\frac{3}{1+ln(1+r-r_0)}-4\Biggr\}\Bigg].
\end{eqnarray}
Therefore from (\ref{eq19}) we obtain the expression of $\rho$ for this model, 
\begin{eqnarray}
\label{eq26}\nonumber
\rho&=&\frac{\lambda}{2(1+\alpha^2\omega^2+\omega^2+\alpha\omega+2\omega-2\alpha\omega^2)}\Bigg[(1-\alpha\omega-2\omega)
+\Biggl\{(-1+\alpha\omega+\omega)^2\\\nonumber
&~&-\frac{(1+\alpha^2\omega^2+\omega^2+\alpha\omega+2\omega-2\alpha\omega^2)ln(1+r-r_0)}{\pi\lambda(1+ln(1+r-r_0))}\Bigg[\frac{\beta}{r^2}-\frac{\beta^2}{r^2}+\frac{(1-r_0)+(1+r-r_0)ln(1+r-r_0)}{r(1+r-r_0)(1+ln(1+r-r_0))ln(1+r-r_0)}\\
&~&+\frac{\beta(1+ln(1+r-r_0))}{2r^2ln(1+r-r_0)}\Biggl\{\frac{(1-r_0)+(1+r-r_0)ln(1+r-r_0)}{(1+ln(1+r-r_0))^2(1+r-r_0)}+\frac{3}{1+ln(1+r-r_0)}-4\Biggr\}\Bigg]\Biggl\}^{\frac{1}{2}}\Bigg].
%p_t&=&\omega\rho,\\
%p_r&=&\alpha\omega\rho.
\end{eqnarray}
\begin{figure}[h]
	\centering
	% \hspace*{\fill}
	\begin{minipage}{.55\textwidth}
		\centering
		\includegraphics[width=.6\linewidth]{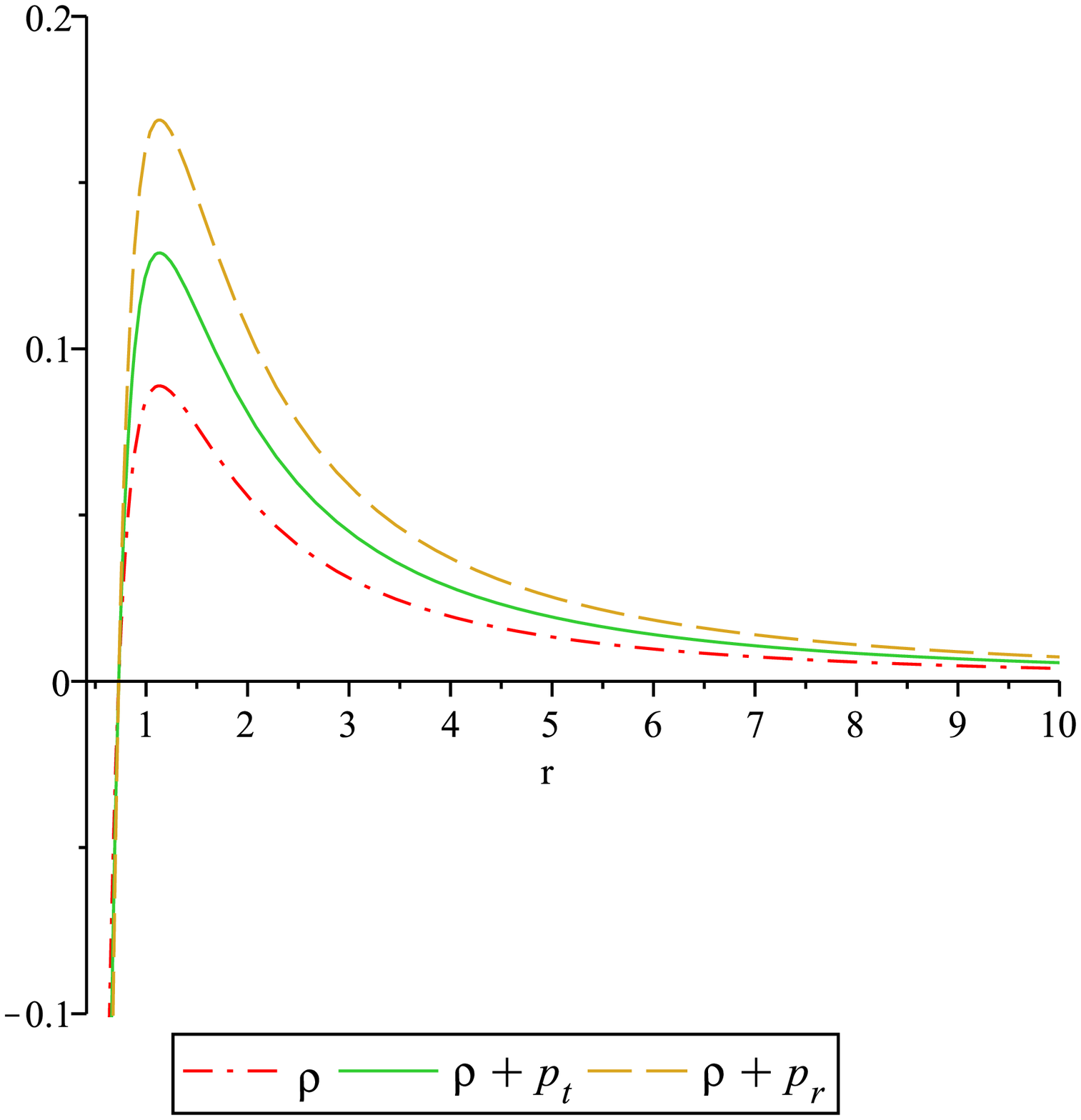}
		\centering FIG.11(A)
	\end{minipage}%%
	\begin{minipage}{.55\textwidth}
		\centering
		\includegraphics[width=.6\linewidth]{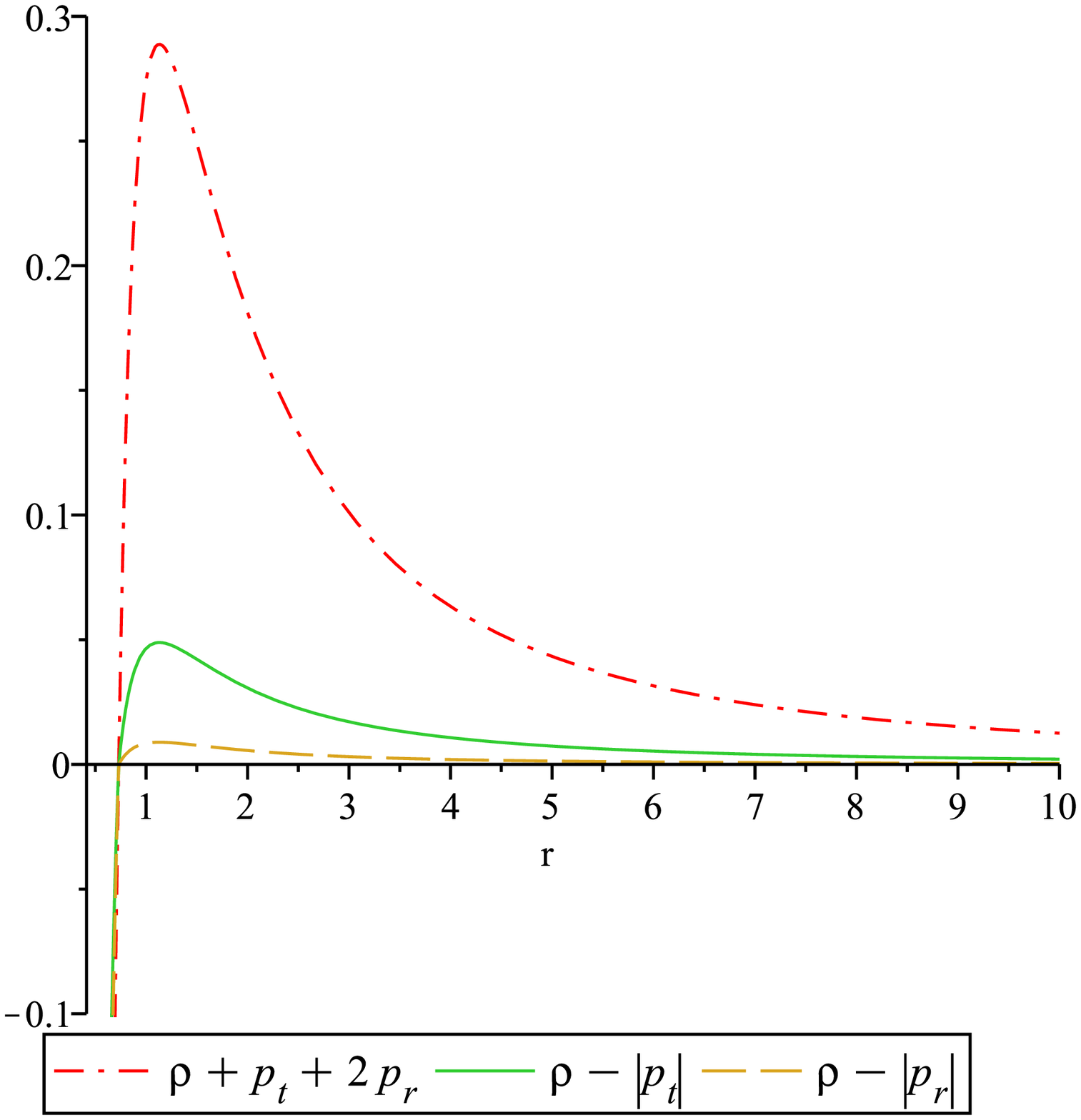}
		\centering {FIG.11(B)}
	\end{minipage}
	
	\caption{Behavior of $\rho$, $\rho+p_t$ and $\rho+p_r$ (FIG.11(A)) and $\rho+p_r+2p_t$, $\rho-|p_t|$ and $\rho-|p_r|$ diagrams (FIG.11(B)) have been plotted for shape function(2) with non-zero tidal force against $r$ when $\alpha=2$, $\omega=0.45$, $\lambda=10^3$, $\beta=-3$ and $r_0=0.5$.}
	\label{fig11}
\end{figure}
\subsubsection{shape function $b(r)=\frac{r}{1+r-r_0},~0<r_0<1$}
For shape function (3), from the equation (\ref{eq11.1}) we get the result,
\begin{eqnarray}\label{eq28}
R&=&\frac{2(r-r_0)}{1+r-r_0}\Biggl[\frac{\beta}{r^2}-\frac{\beta^2}{r^2}+\frac{1-r_0}{r^2(1+r-r_0)}
+\frac{\beta(1+r-r_0)}{2r^2(r-r_0)}\Biggl\{\frac{1-r_0}{(1+r-r_0)^2}+\frac{3}{1+r-r_0}-4\Biggr\}\Biggr].
\end{eqnarray}
Therefore from (\ref{eq19}) we obtain the expression of $\rho$ for this model,
\begin{eqnarray}
\label{eq29}\nonumber
\rho&=&\frac{\lambda}{2(1+\alpha^2\omega^2+\omega^2+\alpha\omega+2\omega-2\alpha\omega^2)}\Bigg[(1-\alpha\omega-2\omega)
+\biggl\{(-1+\alpha\omega+2\omega)^2\\\nonumber
&~&-\frac{(1+\alpha^2\omega^2+\omega^2+\alpha\omega+2\omega-2\alpha\omega^2)(r-r_0)}{\pi\lambda(1+r-r_0)}\Biggl[\frac{\beta}{r^2}-\frac{\beta^2}{r^2}+\frac{1-r_0}{r^2(1+r-r_0)}\\
&~&+\frac{\beta(1+r-r_0)}{2r^2(r-r_0)}\Biggl\{\frac{1-r_0}{(1+r-r_0)^2}+\frac{3}{1+r-r_0}-4\Biggr\}\Biggr]\biggr\}^\frac{1}{2}\Bigg].
%p_t&=&\omega\rho,\\
%p_r&=&\alpha\omega\rho.
\end{eqnarray}
\begin{figure}[h]
	\centering
	% \hspace*{\fill}
	\begin{minipage}{.55\textwidth}
		\centering
		\includegraphics[width=.6\linewidth]{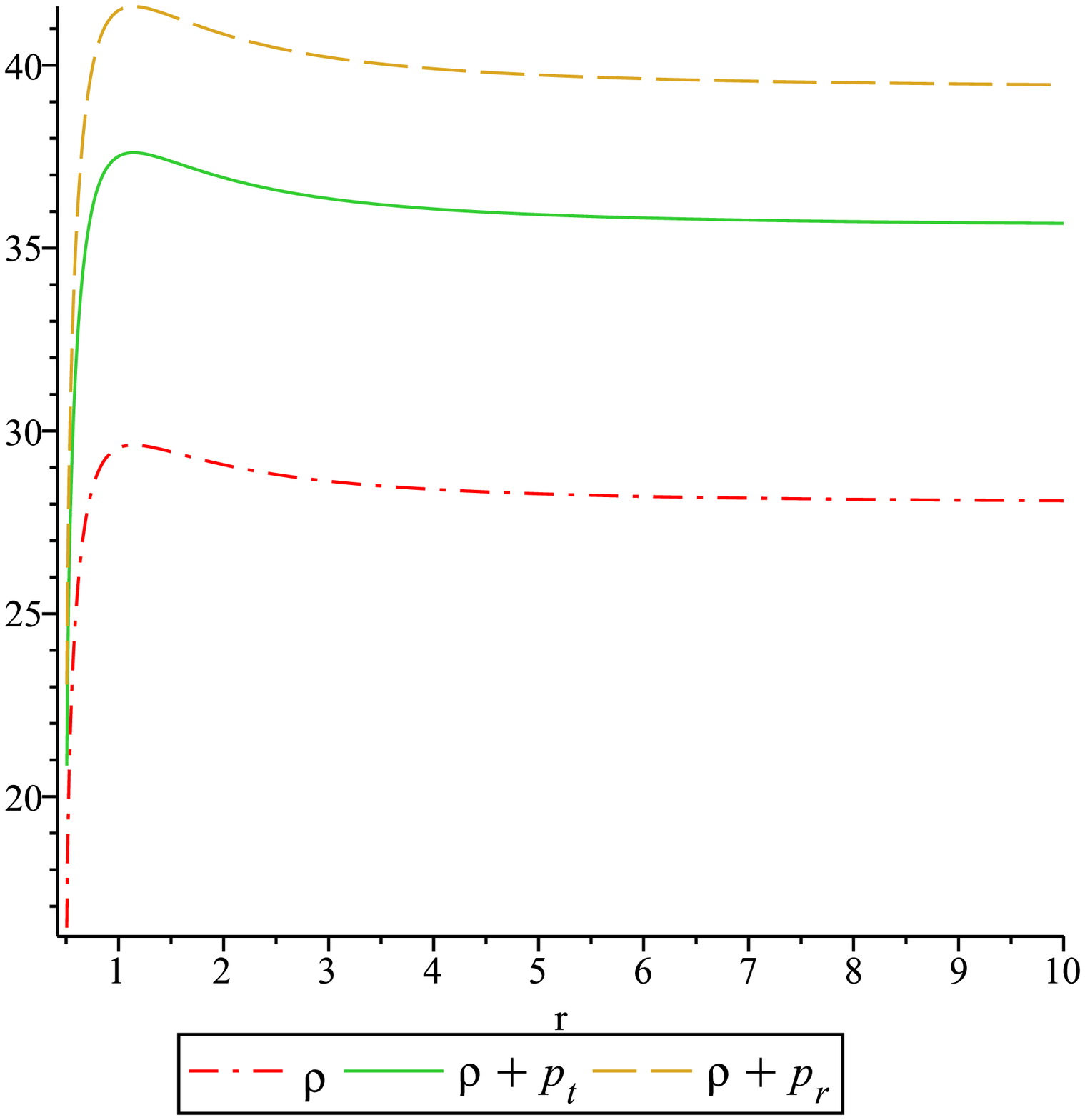}
		\centering FIG.12(A)
	\end{minipage}%%
	\begin{minipage}{.55\textwidth}
		\centering
		\includegraphics[width=.6\linewidth]{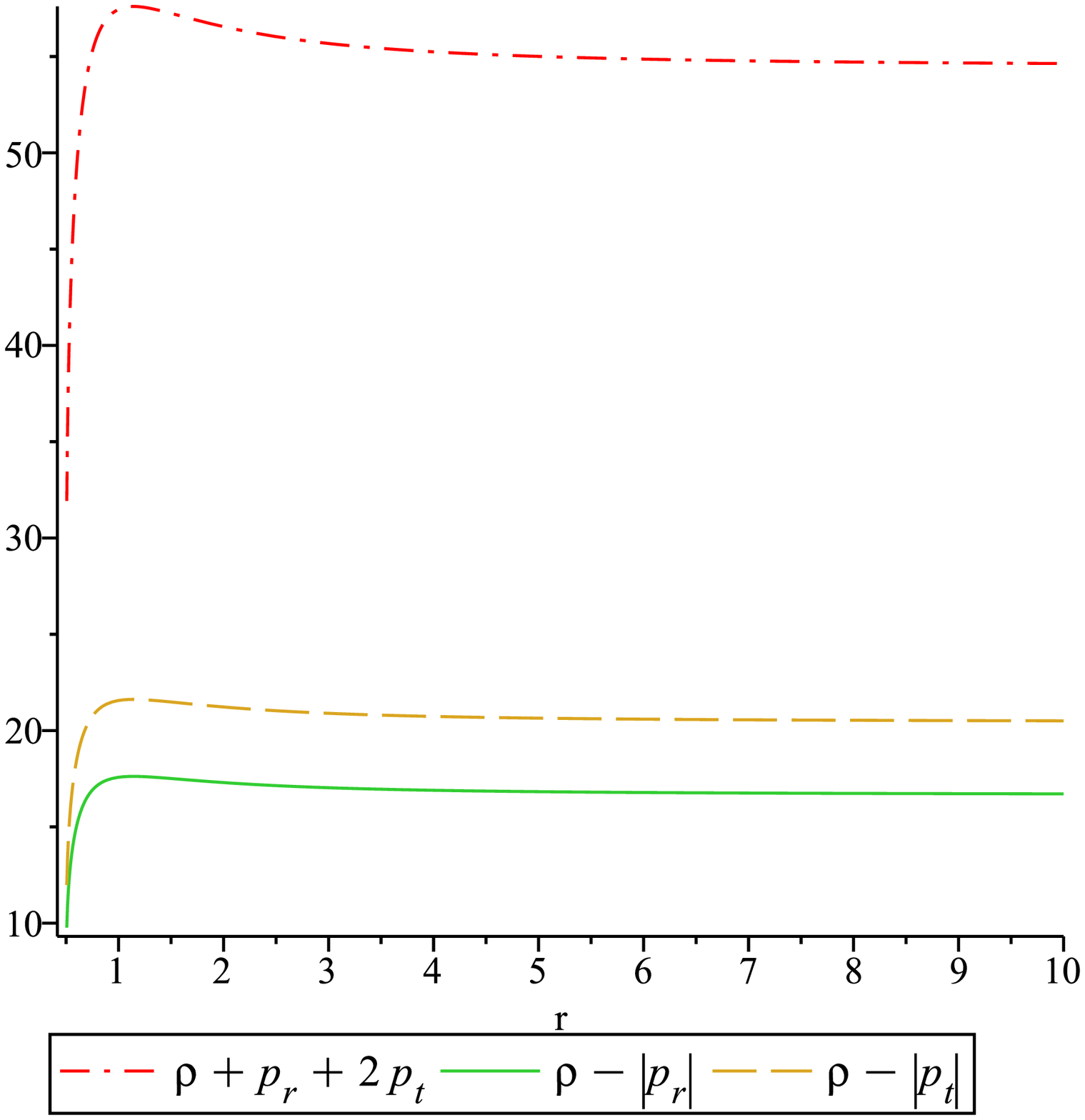}
		\centering FIG.12(B)
	\end{minipage}
	
	\caption{Behavior of $\rho$, $\rho+p_t$ and $\rho+p_r$ (FIG.12(A)) and $\rho+p_r+2p_t$, $\rho-|p_t|$ and $\rho-|p_r|$ diagrams (FIG.12(B)) have been plotted for shape function(3) with non-zero tidal force against $r$ for when $\alpha=1.2$, $\omega=0.27$, $\lambda=10^3$, $\beta=-3$ and $r_0=0.5$.}
	\label{fig12}
\end{figure}
\subsubsection{shape function $b(r)=r_0\biggl\{1+\gamma^2\biggl(1-\frac{r_0}{r}\biggr)\biggr\}$, with $\gamma^2\in(0,1)$}
For shape function (4), from the equation (\ref{eq11.1}) we get the result,
\begin{equation}\label{eq31}
R=-\frac{2}{r}\biggl[\bigl(-\frac{\beta}{r^2}+\frac{\beta^2}{r^2}\bigr)\biggl\{r-r_0\Bigl\{1+\gamma^2\biggl(1-\frac{r_0}{r}\biggr)\Bigr\}\biggr\}-\frac{\gamma^2r_0^2}{r^3}-\frac{\beta}{2r^2}\biggl\{\frac{\gamma^2r_0^2}{r}+3r_0\Bigl\{1+\gamma^2\Bigl(1-\frac{r_0}{r}\Bigr)\Bigr\}-4r\biggr\}\biggr].
\end{equation}
Therefore from (\ref{eq19}) we obtain the expression of $\rho$ for this model,
\begin{eqnarray}
\label{eq32}\nonumber
\rho&=&\frac{\lambda}{2(1+\alpha^2\omega^2+\omega^2+\alpha\omega+2\omega-2\alpha\omega^2)}\Bigg[(1-\alpha\omega-2\omega)
+\biggl\{(-1+\alpha\omega+2\omega)^2\\\nonumber
&~&+\frac{\gamma^2(1+\alpha^2\omega^2+\omega^2+\alpha\omega+2\omega-2\alpha\omega^2)}{r\pi\lambda}\biggl[\bigl(-\frac{\beta}{r^2}+\frac{\beta^2}{r^2}\bigr)\biggl\{r-r_0\Bigl\{1+\gamma^2\biggl(1-\frac{r_0}{r}\biggr)\Bigr\}\biggr\}-\frac{\gamma^2r_0^2}{r^3}\\
&~&-\frac{\beta}{2r^2}\biggl\{\frac{\gamma^2r_0^2}{r}+3r_0\Bigl\{1+\gamma^2\Bigl(1-\frac{r_0}{r}\Bigr)\Bigr\}-4r\biggr\}\biggr]\biggr\}^\frac{1}{2}\Bigg].
%p_t&=&\omega\rho,\\
%p_r&=&\alpha\omega\rho.
\end{eqnarray}
\begin{figure}[h]
	\centering
	% \hspace*{\fill}
	\begin{minipage}{.5\textwidth}
		\centering
		\includegraphics[width=.6\linewidth]{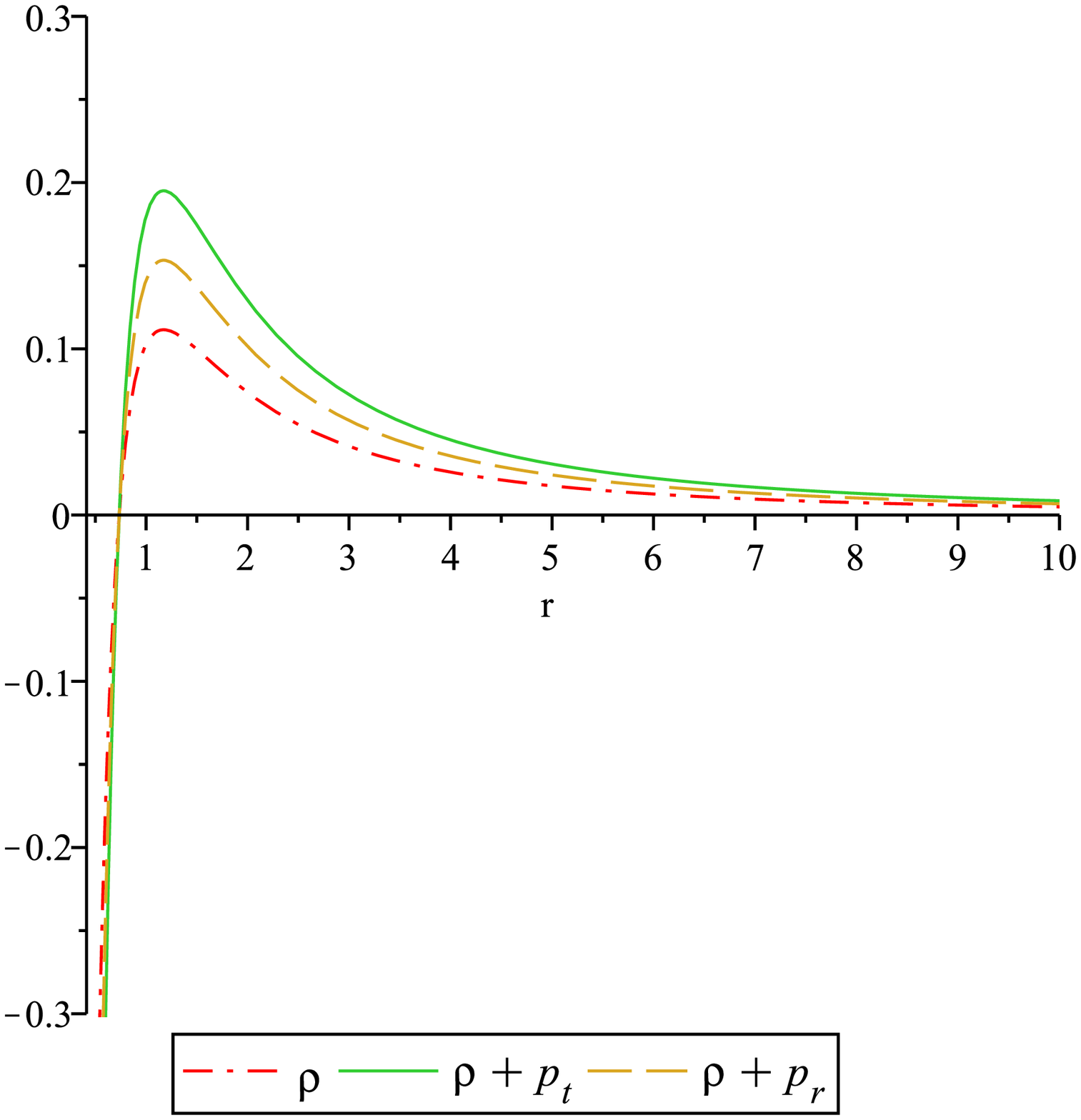}
		\centering FIG.13(A)
	\end{minipage}%%
	\begin{minipage}{.5\textwidth}
		\centering
		\includegraphics[width=.6\linewidth]{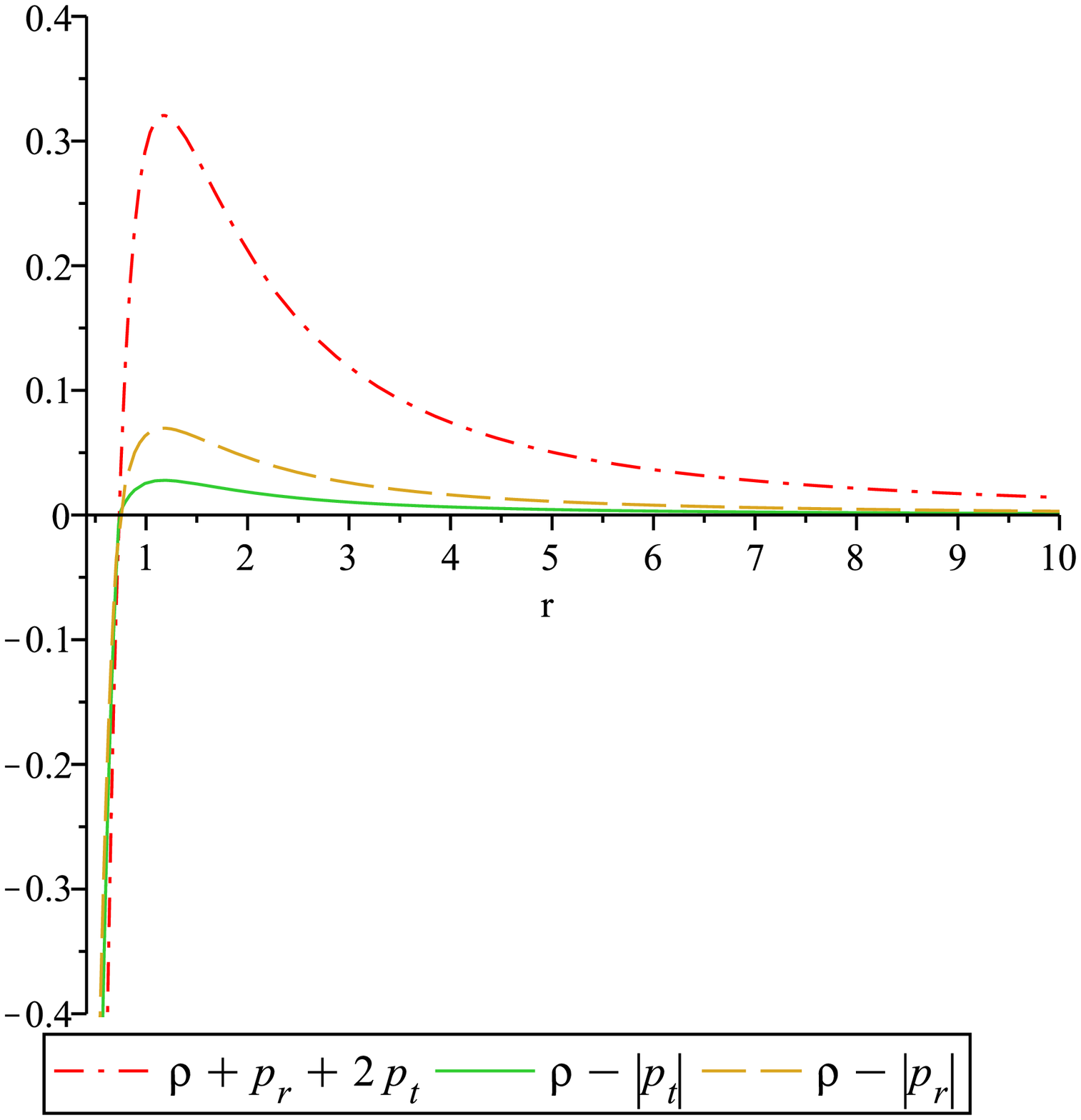}
		\centering FIG.13(B)
	\end{minipage}
	
	\caption{Behavior of $\rho$, $\rho+p_t$ and $\rho+p_r$ (FIG.13(A)) and $\rho+p_r+2p_t$, $\rho-|p_t|$ and $\rho-|p_r|$ diagrams (FIG.13(B)) have been plotted for shape function(4) with non-zero tidal force against $r$ when $\alpha=0.5$, $\omega=0.75$, $\lambda=10^3$, $\gamma=0.8$, $\beta=-3$ and $r_0=0.5$.}
	\label{fig13}
\end{figure}
\subsubsection{ shape function $b(r)=re^{\frac{2\sigma}{\delta}(r^\delta-r_0^\delta)}$ for some constant $\sigma<0$ and $\delta>0$}
For shape function (5), from the equation (\ref{eq11.1}) we get the result,
\begin{equation}\label{eq34}
R=-\frac{2}{r^2}\Bigg[(\beta^2+\beta)-e^{\frac{2\sigma}{\delta}(r^\delta-r_0^\delta)}\{\beta+\beta^2+1+\sigma e^\delta(\beta+2)\}\Bigg].
\end{equation}
Therefore from (\ref{eq19}) we obtain the expression of $\rho$ for this model,
\begin{eqnarray}\label{eq35}
\rho &=&\frac{\lambda}{2(1+\alpha^2\omega^2+\omega^2+\alpha\omega+2\omega-2\alpha\omega^2)}\Bigg[(1-\alpha\omega-2\omega)\\\nonumber
&~&+\sqrt{(-1+\alpha\omega+2\omega)^2+\frac{(1+\alpha^2\omega^2+\omega^2+\alpha\omega+2\omega-2\alpha\omega^2)\biggl\{(\beta^2+\beta)-e^{\frac{2\sigma}{\delta}(r^\delta-r_0^\delta)}\{\beta+\beta^2+1+\sigma e^\delta(\beta+2)\}\biggr\}}{\pi\lambda r^2}}\Bigg].
%p_t&=&\omega\rho,\\
%p_r&=&\alpha\omega\rho.
\end{eqnarray}
\begin{figure}[!]
	\centering
	% \hspace*{\fill}
	\begin{minipage}{.5\textwidth}
		\centering
		\includegraphics[width=.6\linewidth]{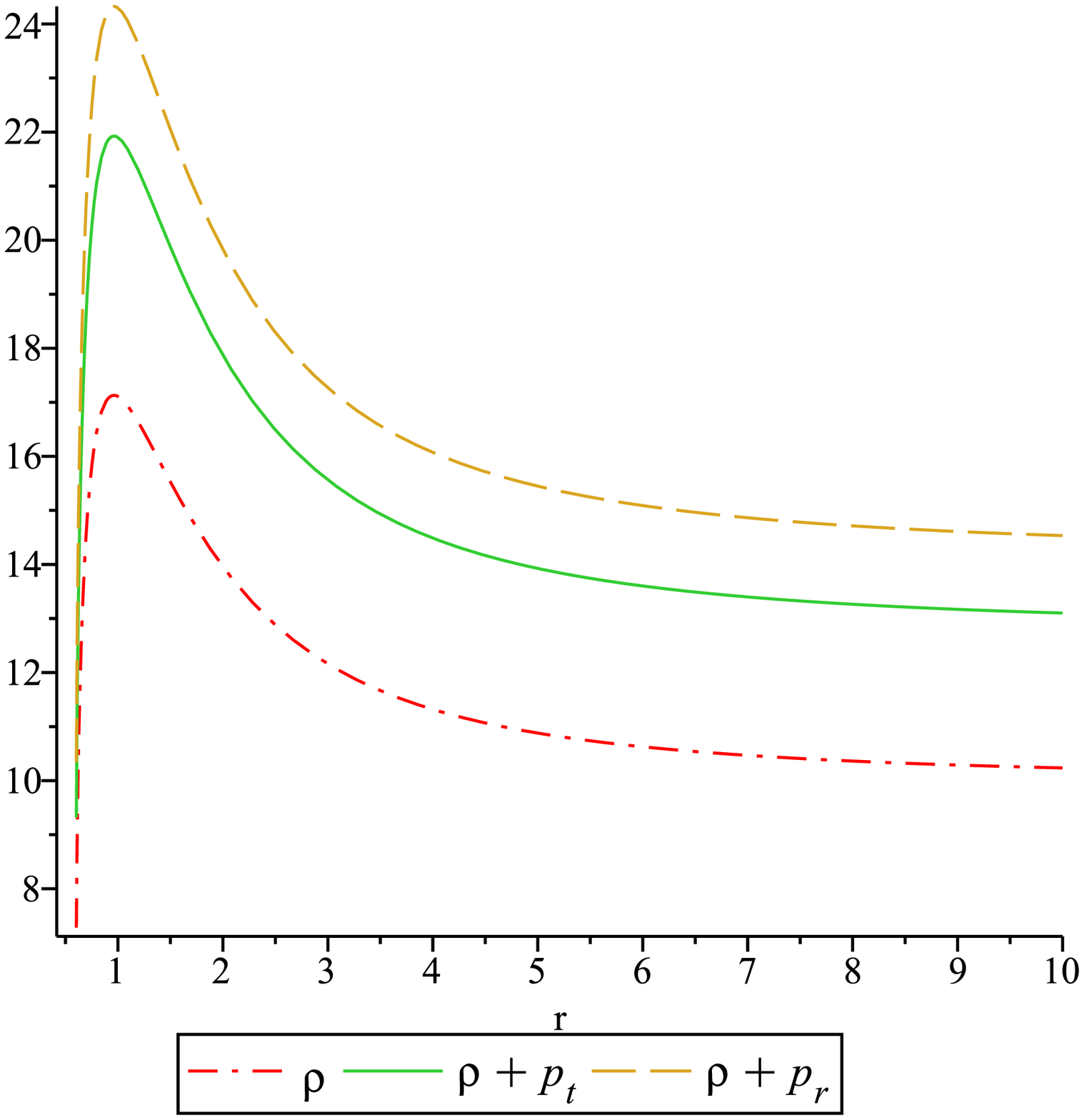}
		\centering FIG.14(A)
	\end{minipage}%%
	\begin{minipage}{.5\textwidth}
		\centering
		\includegraphics[width=.6\linewidth]{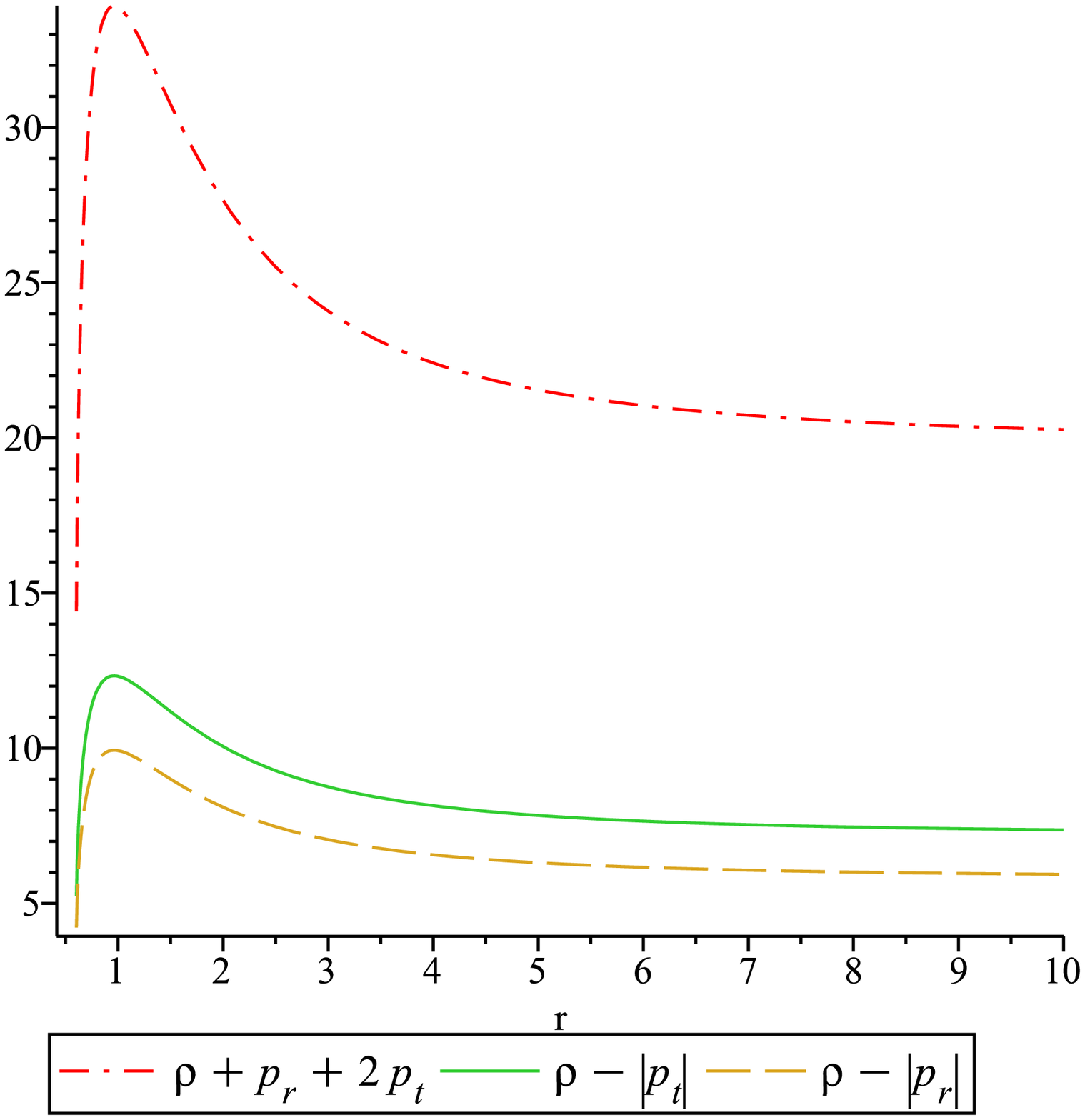}
		\centering FIG.14(B)
	\end{minipage}
	
	\caption{Behavior of $\rho$, $\rho+p_t$ and $\rho+p_r$ (FIG.14(A)) and $\rho+p_r+2p_t$, $\rho-|p_t|$ and $\rho-|p_r|$ diagrams (FIG.14(B)) have been plotted for shape function(5) with non-zero tidal force against $r$  when $\alpha=1.5$, $\omega=0.28$, $\lambda=10^3$, $\sigma=-1$, $\delta=1$, $\beta=-3$ and $r_0=0.5$.}
	\label{fig14}
\end{figure}
\begin{figure}[!]
	\centering
	% \hspace*{\fill}
	\begin{minipage}{.55\textwidth}
		\centering
		
		\includegraphics[width=.6\linewidth]{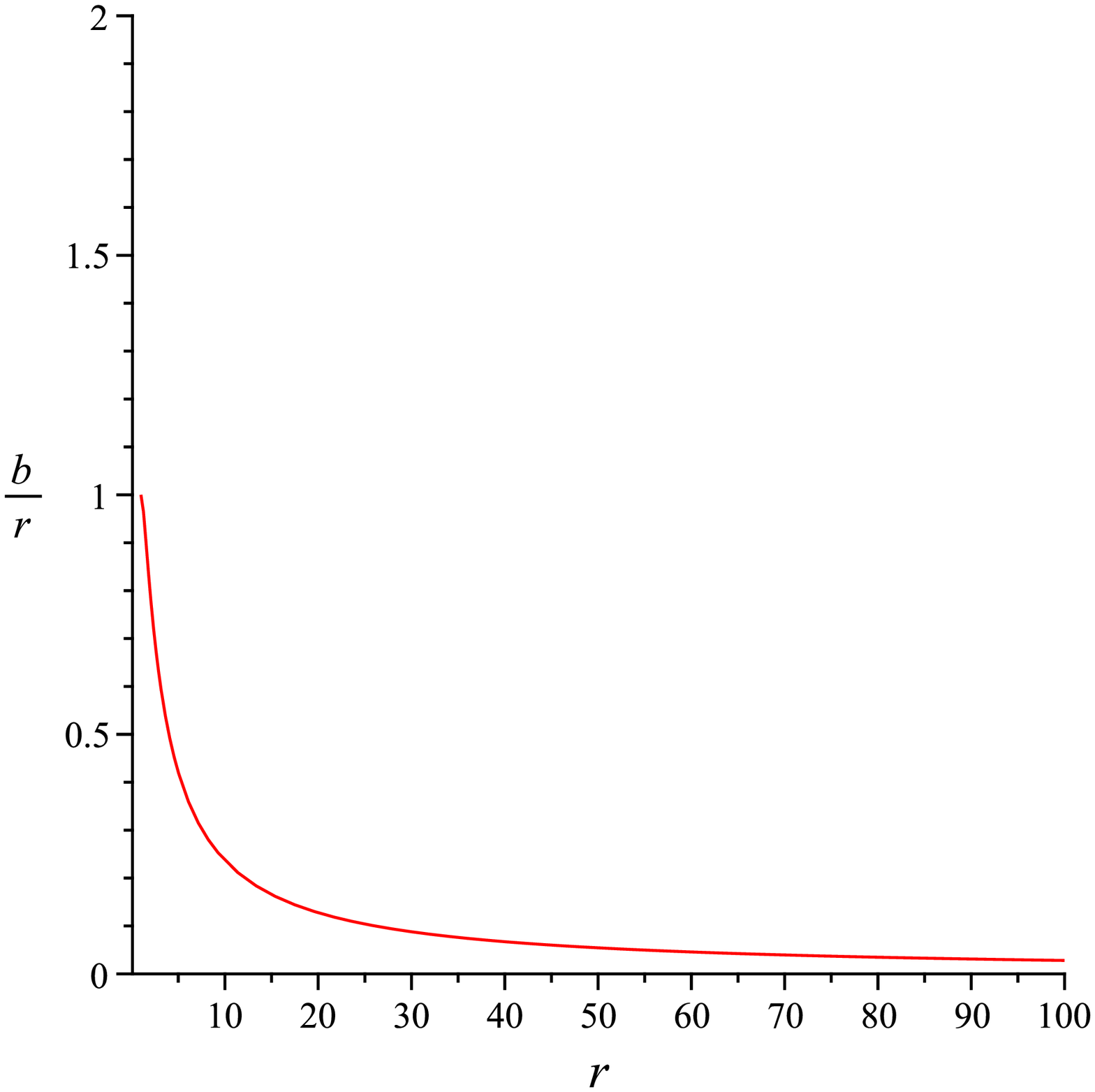}	
		\centering 15(A)
	\end{minipage}%%
	\begin{minipage}{.55\textwidth}
		%\centering
		\includegraphics[width=.6\linewidth]{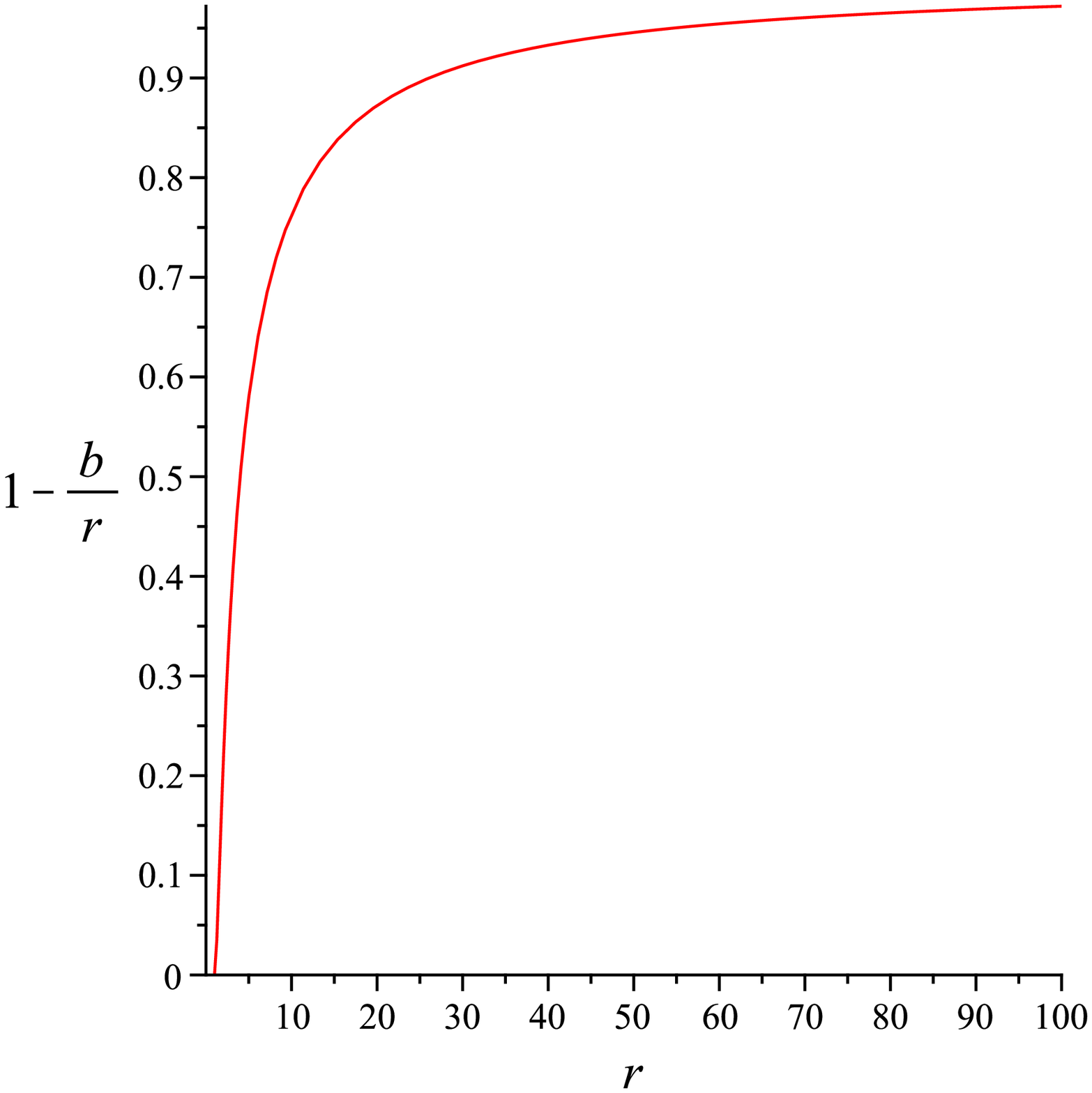}
		\centering 15(B)	
	\end{minipage}
	\caption{Behavior of $\frac{b(r)}{r}$(15(A)), $1-\frac{b(r)}{r}$ (15(B)) have been plotted for obtained shape function(\ref{b}) against $r$ when $n=3.5$, $\lambda=10^4$,$q=1$ and $r_0=1$.}
	\label{fig15}
\end{figure}
\begin{figure}
	\centering
	\includegraphics[width=.6\linewidth]{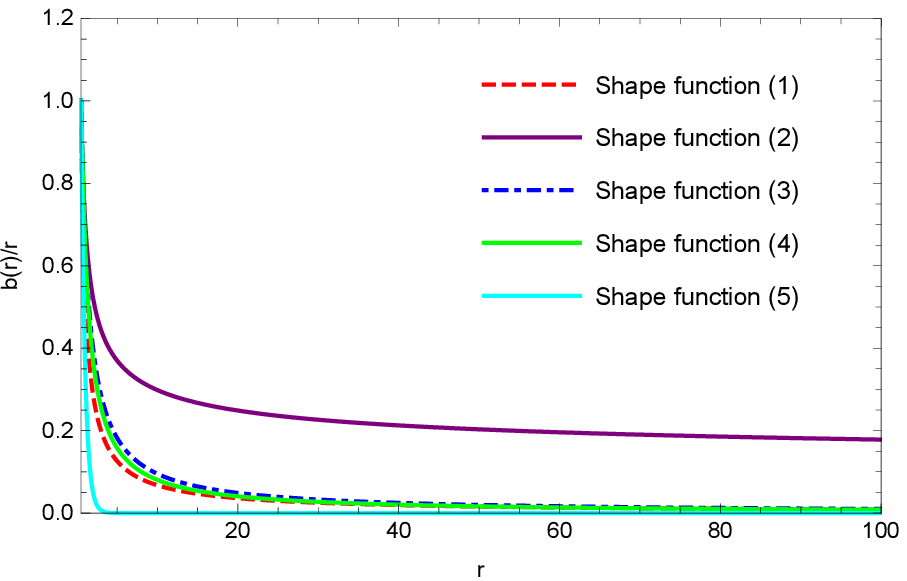}
	\caption{Diagram of $\frac{b(r)}{r}$ vs. ``$r$" for the five different shape functions when $n=0.9$, $\gamma=0.8$, $\sigma=-1$, $\delta=1$ and $r_0=0.5$.}
	\label{fig16}
\end{figure}
\begin{figure}
	\centering
	\includegraphics[width=.6\linewidth]{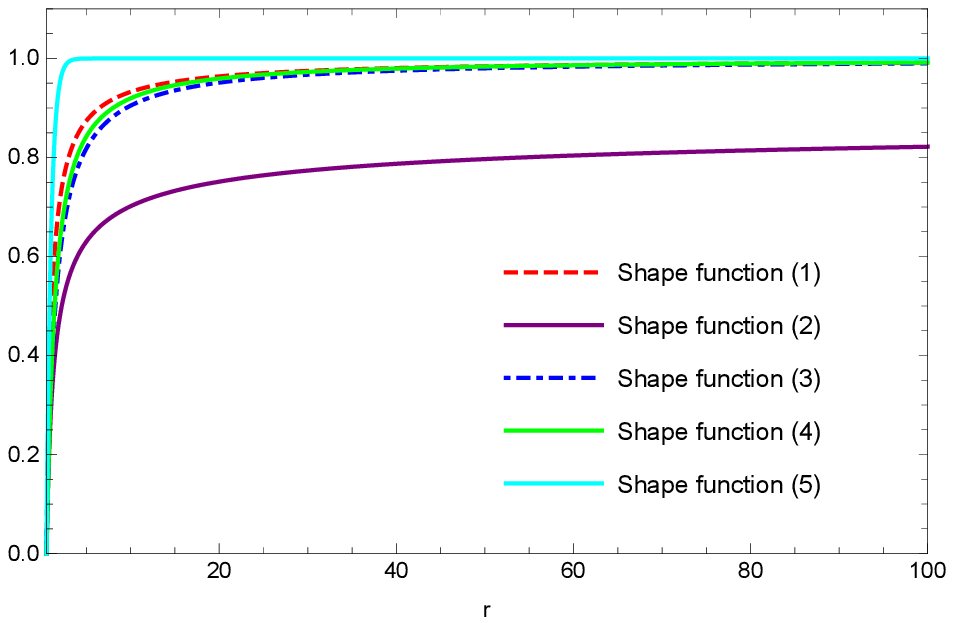}
	\caption{Diagram of $1-\frac{b(r)}{r}$ vs. ``$r$" for the five different shape functions when $n=0.9$, $\gamma=0.8$, $\sigma=-1$, $\delta=1$ and $r_0=0.5$.}
	\label{fig17}
\end{figure}
\begin{figure}[!]
	\centering
	% \hspace*{\fill}
	\begin{minipage}{.55\textwidth}
		\centering
		\includegraphics[width=.6
		\linewidth]{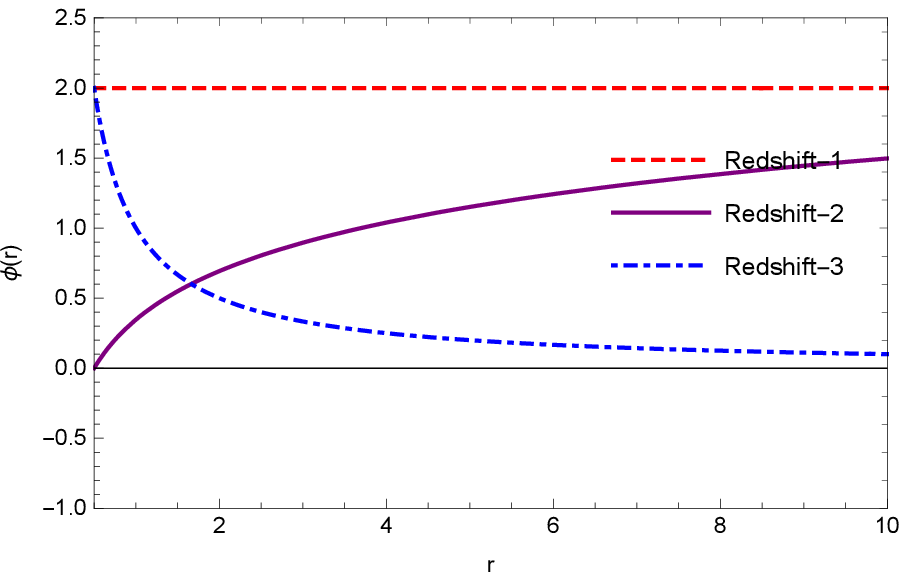}	
		\centering 18(A)
	\end{minipage}%%
	\begin{minipage}{.55\textwidth}
		\centering
		\includegraphics[width=.6\linewidth]{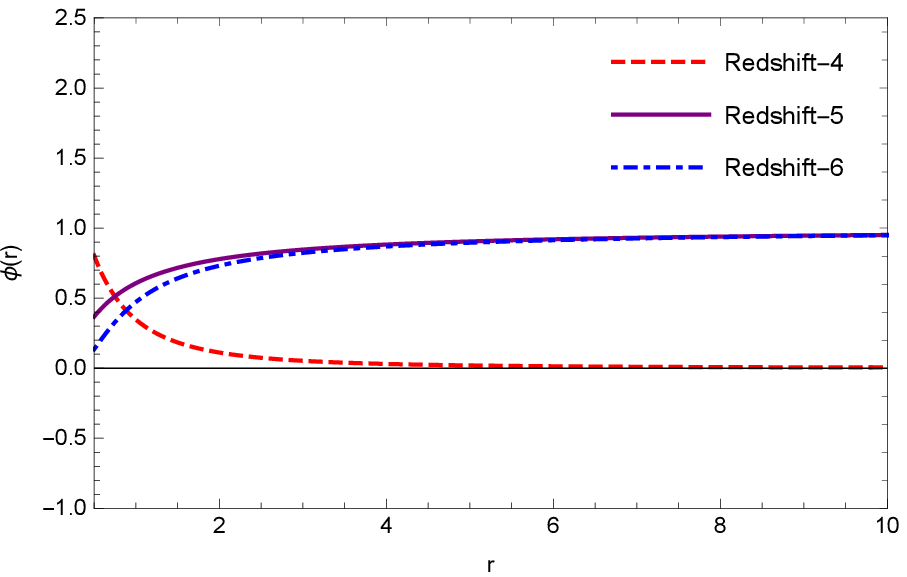}
		\centering 18(B)	
	\end{minipage}
	
	\caption{Diagram of redshift functions  $\phi(r)=$constant(Redshift-1), $\phi(r)=\beta ln\left(\frac{r}{r_0}\right)$(Redshift-2), $\phi(r)=\frac{1}{r}$(Redshift-3), $\phi(r)=ln\frac{\sqrt{\gamma^2+r^2}}{r}$(Redshift-4), $\phi(r)=e^{-\frac{r_0}{r}}$(Redshift-5), $\phi(r)=e^{-\frac{r_0}{r}-\frac{r_0^2}{r^2}}$(Redshift-6) when $\beta=0.5$,$\gamma=1$ and $r_0=0.5$.}
	\label{fig18}
\end{figure}

\section{Results and Discussion}\label{secv}
\begin{table}[!]
	\centering
	\caption{Range of `$r$' where energy conditions are satisfied for obtained shape function(\ref{b}) when $n\in(3, 5)$ under isotropic scenario.}
	\begin{tabular}{|>{\bfseries}c|*{5}{c|}}\hline
		{\bfseries Redshift functions}  & {NEC} &{WEC} & {SEC} & {DEC} 
		\\ \hline
		\text{ $\phi(r)=\text{constant}$}         &  \text{$(r_0, 2)$} &\text{$(r_0, 2)$} & \text{$\times$}& \text{$(r_0, 2)$} \\ \hline
		\text{ $\phi(r)=\beta ln\left(\frac{r}{r_0}\right)$} &  \text{$(r_0, 5)$} &  \text{$(r_0, 5)$} & \text{$(r_0, 5)$} &\text{$(r_0, 1.8)$}
		\\ \hline
		\text{$\phi(r)=\frac{1}{r}$} &\text{$~(r_0, 1.56)~$} & \text{$~(r_0, 1.56)~$} &\text{$\times$} &\text{$~(r_0, 1.56)~$}
		\\ 
		\hline
		\text{$\phi(r)=ln\left(\frac{\sqrt{\gamma^2+r^2}}{r}\right)$}	&\text{$(r_0, 2)$} & \text{$(r_0, 2)$} &\text{$\times$}&\text{$(r_0, 2)$}
		\\ 
		\hline
		\text{$\phi(r)=e^{-\frac{r_0}{r}}$} &\text{$(r_0, 5)$} & \text{$(r_0, 5)$} 	&\text{$(r_0, 5)$} & \text{$(r_0, 2)$}
		\\ 
		\hline	
		\text{$\phi(r)=e^{-\frac{r_0}{r}-\frac{r_0^2}{r^2}}$} &\text{$(r_0, 5)$} & \text{$(r_0, 5)$} 	&\text{$(r_0, 5)$}&\text{$~(r_0, 1.75)~$}
		\\ 
		\hline
	\end{tabular}
	\label{Table:T1}
\end{table}
A new shape function is obtained by considering $\rho=q\left({\frac{r}{r_0}}\right)^{-n}$ and $\epsilon=0$. Now the case $\epsilon=0$ can be obtained when the space is conformally flat or in the vacuum space. Also it is observed that when $\epsilon=0$ then the presence of exotic matter may depend upon the sign of $\sigma_r$. In other words if $\sigma_r<0$ then the presence of exotic matter will be ensured (when $\rho$ and $\lambda$ both are positive)\cite{r35}. Wormhole solutions under many redshift functions are also obtained in this paper. In figure (\ref{fig4}), (\ref{fig6}), (\ref{fig7}), since $p$ is negative for these redshift functions so the figure of $\rho+p$ and $\rho-|p|$ will be identical for each case. All the energy conditions are examined in (FIG.\ref{fig3}---FIG.\ref{fig9}) for each wormhole solutions corresponding to mentioned redshift functions (for isotropic scenario). It is seen that for redshift functions $\phi(r)= \text{constant}$, $\phi(r)=\frac{1}{r}$, $\phi(r)=ln\frac{\sqrt{\gamma^2+r^2}}{r}$ there exists a region for each cases where all energy conditions are satisfied except SEC for $n\in(3,5)$. For the redshift functions $\phi(r)=\beta ln\left(\frac{r}{r_0}\right)$, $\phi(r)=e^{-\frac{r_0}{r}}$ and $\phi(r)=e^{-\frac{r_0}{r}-\frac{r_0^2}{r^2}}$ , all energy conditions are satisfied in a neighbourhood of the wormhole throat for $n\in(3,5)$ , which is clear from table \ref{Table:T1}.
\par 
For anisotrpic case, from the energy conditions, it can be observed from inequalities (\ref{eq13.1})--(\ref{eq16.1}), if the relation between tangential and radial pressure is of the from $p_t=\omega\rho$ and $p_r=\alpha p_t$ ($\alpha\neq1$), then all energy conditions will satisfy if $|\alpha\omega|\leq1$, $|\omega|\leq1$ and $\omega(\alpha+2)\geq-1$ provided $\rho\geq0$. Thus if $\alpha\omega$, $\omega$ $\in(0, 1)$, then all energy conditions will be satisfied in the region where $\rho\geq0$. Hence from the FIG.\ref{fig10}--FIG.\ref{fig14}, the region where $\rho\geq0$, all energy conditions will satisfy.
\par 
For redshift function $\phi(r)=\beta ln\left(\frac{r}{r_0}\right)$, the connecting asymptotic spaces by the wormhole are not flat in nature, in other all the considered cases connecting asymptotic spaces are flat except for the shape function (2)(from figures \ref{fig15}A, \ref{fig16} and \ref{fig18}).
From figures \ref{fig15}(B) and \ref{fig17}, we can conclude that all the wormholes corresponding to all shape functions throughout in this work are infinitely extendable\cite{r36}.
 
\par 
Hence violation of NEC in the neighbourhood of wormhole throat which is necessary in Einstein gravity\cite{r6},\cite{r7},\cite{r4} to form a wormhole, is not necessary in brane-world gravity theory.
% More-over it is observed that, in maximum cases all energy conditions are satisfied in wormhole-throats neighbourhood. 

\section*{Acknowledgement}
We thank Prof. Subenoy Chakraborty, Jadavpur University for the useful discussions about this work.

\end{document}